\documentclass[10pt]{elsarticle}
\usepackage{amssymb,amsmath}
\usepackage{enumitem}
\usepackage{bbm}
\usepackage{upgreek}
\usepackage{mathtools}
\usepackage[margin=1in,papersize={8.5in,11in}]{geometry}
\usepackage[utf8]{inputenc}
\usepackage[english]{babel}
\usepackage{amsthm}
\usepackage{float}
\usepackage{amsfonts,mathrsfs}
\usepackage{bm}
\usepackage{makecell}

\usepackage{tikz}
\usetikzlibrary{matrix}

\usepackage{epstopdf}
\usepackage{subcaption}
\usepackage{tablefootnote}
\graphicspath{ {images/} }
\usepackage{comment}

\newtheorem{prop}{Proposition}
\theoremstyle{definition}

\def\R{\mathbb{R}}
\def\H{\mathcal{H}}

\def\S{\mathcal{S}} 
\def\L{\mathcal{L}} 
\def\M{\mathcal{M}}

\def\P{\mathcal{P}}
\def\U{\mathcal{U}}

\def\Q{\mathcal{Q}}

\def\G{\mathcal{G}}
\def\K{\mathcal{K}} % Koopman
\def\I{\mathcal{I}}
\def\U{\mathcal{U}}

\def\Pi{\mathbf{P}}

\def\tP{\tilde{\mathcal{P}}}
\def\tQ{\tilde{\mathcal{Q}}}
\def\tK{\tilde{\mathcal{K}}}
\journal{arXiv}

\begin{document}
\author[ucm]{Yuanran Zhu\corref{correspondingAuthor}}
\cortext[correspondingAuthor]{Corresponding author}
\ead{yzhu56@ucmerced.edu}
\author[msu]{Huan Lei}
\address[ucm]{Department of Applied Mathematics, University of California, Merced\\ Merced (CA) 95343}
\address[msu]{Department of Computational Mathematics, Michigan State University\\ East Lansing (MI) 48824}

% \author{Yuanran Zhu}
% \address[Yuanran Zhu]{Department of Applied Mathematics, University of California, Merced\\ Merced (CA) 95343, USA}
% \email[Corresponding author]{yzhu56@ucmerced.edu}
% \author{Huan Lei}
% \address[Huan Lei]{Department of Computational Mathematics, Science \& Engineering and Department of Statistics \& Probability, Michigan State  University, MI 48824, USA}
% \email{leihuan@msu.edu}

\begin{abstract}
Built upon the hypoelliptic analysis of the effective Mori-Zwanzig (EMZ) equation for observables of stochastic dynamical systems, we show that the obtained semigroup estimates for the EMZ equation can be used to derive prior estimates of the observable statistics for system in the equilibrium and nonequilibrium state. In addition, we introduce both first-principle and data-driven methods to approximate the EMZ memory kernel, and prove the convergence of the data-driven parametrization schemes using the regularity estimate of the memory kernel. The analysis results are validated numerically via the Monte-Carlo simulation of the Langevin dynamics for a Fermi-Pasta-Ulam chain model. With the same example, we also show the effectiveness of the proposed memory kernel approximation methods.



\end{abstract}

\title{Effective Mori-Zwanzig equation for the reduced-order modeling of stochastic systems}
\maketitle

\section{Introduction}
The projection operator method, which is also known as the Mori-Zwanzig (MZ) formulation \cite{Mori,zwanzig1961memory}, is a widely used dimension-reduction framework in statistical mechanics.  
The key feature of such formulation 
is that it allows us to formally derive the
generalized Langevin equations (GLEs) 
\cite{zwanzig1973nonlinear,Chorin,Venturi_PRS,Grogan_Lei_JCP_2020} for 
coarse-grained quantities of interest based on microscopic equations of motion. Such GLEs can be found in a variety of 
applications, including molecular dynamics 
\cite{Li2015,Yoshimoto2013,
espanol1995statistical,espanol1995hydrodynamics}, 
fluid mechanics \cite{parish2017non,Falkena2019}, 
and, more generally, systems described by nonlinear 
partial differential equations (PDEs)
\cite{venturi2014convolutionless,stinis2004stochastic,
lu2017data,lin2019data}. 
Although being used in the physics and applied mathematics communities for a rather long time, a systematic study of the MZ equation within an rigorous analytical framework is still lacking. This is closely related to the well-known difficulty on the quantification of the orthogonal dynamics in the MZ equation. Being a high-dimensional flow 
which is generated by an integro-differential operator, the mathematical properties such as the regularity and ergodicity of the orthogonal dynamics are not well understood. Hence from a theoretical point of view, there is no available prior estimate which helps to determine the properties of the MZ memory integral and the fluctuation force. As a result, the numerical approximations of these terms has to be done in an rather {\em ad hoc} manner.

Some recent works have shed light on this direction. In particular, Kupferman, Givon
and Hald proved \cite{givon2005existence} the existence and uniqueness of the orthogonal dynamics for a classical dynamical system
with Mori's projection operator. More recently, Zhu and Venturi \cite{zhu2018estimation} were able to get the uniform boundedness of the orthogonal
dynamics propagator for Hamiltonian systems using semigroup estimates \cite{zhu2018estimation}. The theoretical result obtained therein was later extended and greatly improved for the analysis of the effective Mori-Zwanzig (EMZ) equation corresponding to stochastic differential equations (SDEs) \cite{zhu2020hypoellipticity}. In particular, they developed a 
thorough mathematical analysis of the EMZ equation using the hypoelliptic technique developed mainly by H\'erau, Nier, Eckmann, Hairer and Helffer \cite{herau2004isotropic,eckmann2003spectral,nier2005hypoelliptic}. The key finding is that the ergodicity and regularity of the stochastic flow generated by the Markovian semigroup $e^{-t\K}$, where $\K$ is the Kolmogorov operator corresponding to the SDE, implies the ergodicity and regularity of the stochastic flow generated by the EMZ {\em orthogonal} semigroup $e^{-t\Q\K\Q}$, provided that $\P=\I-\Q$ is a Mori-type projection operator. This connection enables us to get a clear understanding on the dynamical properties of the orthogonal dynamics generated by $e^{-t\Q\K\Q}$.

In this work, we continue Zhu and Venturi's hypoelliptic study of the EMZ equation for stochastic dynamical systems. The main objective of the paper is twofold. First, we apply the semigroup estimate obtained in \cite{zhu2020hypoellipticity} to different stochastic systems and show that it enables us to derive useful prior estimates for the statistics of observables. In particular, we prove that the reduced-order observables in some commonly used stochastic models have exponentially decaying time autocorrelation function and EMZ memory kernel. This fact verifies the frequently used exponentially decaying assumption for the memory kernel from a theoretical point of view. Secondly, we will demonstrate the effectiveness of the series expansion approximation method for the memory kernel reconstruction of the EMZ equation. To this end, we will focus on the first-principle parametrization method \cite{zhu2019generalized} and the data-driven methods \cite{baczewski2013numerical,berkowitz1981memory,chu2017mori,Li2017} developed over the years. For the numerical examples we considered, these two methods are proven to yield accurate simulation result within the range of their applicability. Moreover, we will prove the convergence of the commonly used data-driven method using the regularity estimate for the orthogonal dynamics. For the reduced-order modeling problem of a large-scale stochastic system, the proposed analysis for the EMZ equation shows the potential usage of the hypoelliptic method in analyzing the dynamical behavior of the reduced-order model. The numerical methodology provides a practical way to solve it. 

This paper is organized as follows. Section \ref{sec:EMZE} briefly reviews the derivation of the effective Mori Zwanzig (EMZ) equation for the stochastic dynamical system driven by white noise. In Section \ref{sec:Ana}, we focus on the equilibrium and nonequilibrium dynamics of the interacting anharmonic chains and derive prior estimates for various observable statistics such as the time autocorrelation function, the nonequilibrium mean, the EMZ memory kernel and the fluctuation force. In Section \ref{sec:param_memory}, we introduce different parametrization methods to approximate the EMZ memory kernel and prove their  convergence. All these theoretical results are verified numerically in Section \ref{sec:app} via the simulation of the Langevin dynamics for a Fermi-Pasta-Ulam chain model. The main findings of this paper are summarized in Section \ref{sec:conclusion}.

\section{Effective Mori-Zwanzig equation for stochastic system}
\label{sec:EMZE}
The starting point of our work is the Mori-Zwanzig equation for the stochastic dynamical systems. Such a equation has been derived by different researchers \cite{morita1980contraction,espanol1995hydrodynamics,hudson2018coarse,zhu2020hypoellipticity}. Here we adopt the formulation introduced in \cite{zhu2020hypoellipticity}. To this end, we consider a $d$-dimensional stochastic 
differential equation in $\R^d$:
\begin{align}\label{eqn:sde}
\frac{d\bm x(t)}{dt}=\bm F(\bm x(t))+\bm \sigma(\bm x(t))\bm \xi(t), \qquad \bm x(0)=\bm x_0\sim \rho_0(\bm x),
\end{align}
where $\bm F:\R^d \rightarrow \R^d$ and 
$\bm \sigma: \R^d\rightarrow \R^{d\times m}$ are 
smooth functions. $\bm \xi(t)$ is a
$m$-dimensional Gaussian white noise with 
independent components, and $\bm x_0=\bm x(0)$ is a 
random initial state characterized in terms of 
a probability density function $\rho_0(\bm x)$. It is well known that the system of SDEs \eqref{eqn:sde} induces a $d$-dimensional Markovian process in $\R^d$. This allows us to define a 
composition operator $\M(t,0)$ that pushes forward 
in time the average of the observable $\bm u(t)=\bm u(\bm x(t))$ 
over the noise, i.e., 
\begin{align}
\mathbb{E}_{\bm \xi(t)}[\bm u(\bm x(t))|\bm x_0]= 
\M(t,0)\bm u(\bm x_0)=e^{t\K}\bm u(\bm x_0).
\label{MarkovSemi}
\end{align}
Using It\^o's interpretation for the stochastic integral, we note that $\M(t,0)$ is a Markovian semigroup
generated by the following (backward) Kolmogorov operator 
\cite{Risken,kloeden2013numerical}:
\begin{align}
\K(\bm x_0)&=\sum_{k=1}^dF_k(\bm x_0)\frac{\partial}{\partial x_{0k}}
+\frac{1}{2}\sum_{j=1}^m\sum_{i,k=1}^d\sigma_{ij}(\bm x_0)
\sigma_{kj}(\bm x_0)\frac{\partial}{\partial x_{0i}\partial x_{0k}}.
\label{KI}
%\K^S(\bm x_0)&=\sum_{k=1}^d\left[F_k(\bm x_0)
%-\frac{1}{2}\sum_{j=1}^m\sigma_{kj}(\bm x_0)
%\sum_{i=1}^d\sigma_{ij}(\bm x_0)
%\frac{\partial}{\partial x_{0i}}\right]
%\frac{\partial}{\partial x_{0k}},
%\label{KS}
\end{align}
With the evolution operator $\M(t,0)$ available, we can now derive the Mori-Zwanzig equation for noise-averaged quantity $\mathbb{E}_{\bm \xi(t)}[\bm u(\bm x(t))|\bm x_0]$.
To this end, we introduce a projection operator $\P$ and 
the complementary projection $\Q=\I-\P$. 
By differentiating Dyson's identity 
\cite{zhu2019generalized,zhu2018faber,dominy2017duality} for the Markovian semigroup $\M(t,0)$, we can obtain the exact evolution equation governing 
the evolution of \eqref{MarkovSemi}:
\begin{equation}
\frac{\partial}{\partial t}e^{t\K}\bm u(0)
=e^{t\K}\mathcal{PK}\bm u(0)
+e^{t\Q\K\Q}\mathcal{QK}\bm u(0)+\int_0^te^{s\K}\P\K
e^{(t-s)\Q\K\Q}\mathcal{QK}\bm u(0)ds,\label{eqn:EMZ_full}
\end{equation}
where $\bm u(0)=\bm u(\bm x_0)$. Note that in \eqref{eqn:EMZ_full}, $e^{t\Q\K}\Q\K$ is replaced by a another operator $e^{t\Q\K\Q}\Q\K$ which makes it slightly different from the commonly used MZ equation \cite{hudson2018coarse,espanol1995hydrodynamics}. Such a modification is needed for the semigroup estimation we are going to present. It is possible because $\Q$ is a projection operator, $e^{t\Q\K}$ and  $e^{t\Q\K\Q}$ are equivalent in the range of $\Q$. The three terms on the right hand side of \eqref{eqn:EMZ_full} 
are called streaming term, fluctuation 
(or noise) term, and memory term respectively. It is often 
useful to compute the evolution 
of the observable $\bm u(t)$ within a closed 
linear space such as the image of the projection operator 
$\P$. Hence we apply the projection operator 
$\P$ to \eqref{eqn:EMZ_full} and get the projected equation:
\begin{equation} 
\frac{\partial}{\partial t}\mathcal{P}e^{t\K}\bm u(0)
=\mathcal{P}e^{t\K}\mathcal{PK}\bm u(0)
+\int_0^t\P e^{s\K}\mathcal{PK}
e^{(t-s)\Q\K\Q}\mathcal{QK}\bm u(0)ds.\label{eqn:EMZ_projected}
\end{equation}
Eqn \eqref{eqn:EMZ_full} and its projected form \eqref{eqn:EMZ_projected} only describe the noise-averaged dynamics of the observable $\bm u(\bm x(t))$, hence they are called as the {\em effective} Mori-Zwanzig (EMZ) equations for the stochastic system. The EMZ equation and the classical MZ
equation for deterministic (autonomous) systems \cite{zhu2019generalized,zhu2018estimation,zhu2018faber} have the same structure. The only difference is that the Liouville operator $\L$ is replaced by a Kolmogorov operator $\K$. 

In this paper,
we mainly focus on the EMZ equation corresponding to
Mori-type linear projection operator. To derive such a equation,
we consider the weighted Hilbert space 
$H=L^2(\R^d,\rho)$, where $\rho$ is a positive weight 
function in $\R^d$. Let
\begin{equation}
\langle h,g\rangle_{\rho}=\frac{1}{\int \rho d\bm x}\int 
h(\bm x)g(\bm x)\rho(\bm x)d\bm x 
\qquad h,g\in H
\label{ip}
\end{equation}
be the inner product in $H$. 
The Mori-type projection operator $\P$ is a finite-rank operator in $H$ with the canonical form: 
\begin{align}
\label{Mori_P}
\P h=\sum_{i,j=1}^M G^{-1}_{ij}
\langle u_i(0),h\rangle_{\rho}u_j(0),
\qquad h\in H,
\end{align}
where $G_{ij}=\langle u_i(0),u_j(0)\rangle_{\rho}$
and $u_i(0)=u_i(\bm x(0))$ ($i=1,...,M$) are
$M$ linearly independent functions with respect to inner product $\langle \cdot,\cdot\rangle_{\rho}$. Since $\P$ is a finite rank operator, we can rewrite 
the EMZ equations \eqref{eqn:EMZ_full}-\eqref{eqn:EMZ_projected} equivalently as: 
\begin{align}
\frac{d\bm u(t)}{dt} &= \bm \Omega \bm u(t) +
\int_{0}^{t}\bm K(t-s)\bm u(s)ds+\bm f(t),\label{gle_full}\\
\frac{d}{dt}\P{\bm u}(t) &= \bm \Omega\P {\bm u}(t)+ 
\int_{0}^{t} \bm K(t-s) \P {\bm u}(s)ds,
\label{gle_projected}
\end{align}
where $\bm u(t) =[u_1(t),\dots,u_M(t)]^T$ and 
\begin{subequations}
\begin{align}
		G_{ij} & = \langle u_{i}(0), u_{j}(0)\rangle_{\rho}
		\quad \text{(Gram matrix)},\label{gram}\\
		\Omega_{ij} &= \sum_{k=1}^MG^{-1}_{jk}
		\langle u_{k}(0), \K u_{i}(0)\rangle_{{\rho}}\quad 
		\text{(streaming matrix)},\label{streaming}\\
		K_{ij}(t) & =\sum_{k=1}^M G^{-1}_{jk}
		\langle u_{k}(0), \K e^{t\Q\K\Q}\Q\K u_{i}(0)\rangle_{\rho}\quad 
		\text{(memory kernel)},\label{SFD}\\
		 f_i(t)& =e^{t\Q\K\Q}\Q\K u_i(0) \quad 
		\text{(fluctuation term)}.\label{f}
	\end{align}
\end{subequations}
\noindent
To be noticed that here we allow a slight abuse of notation and use $\bm u(\bm x(t))$ to represent its noise average $\mathbb{E}_{\bm\xi(t)}[\bm u(\bm x(t))|\bm x_0]$. This applies to all EMZ equations in the following sections. In statistical mechanics, the EMZ equation \eqref{gle_full} and \eqref{gle_projected} are often called as the generalized Langevin equations (GLEs). The projection operator method provides a systematic way to derive such {\em closed} equations of motion for reduced-order observables $\bm u(\bm x(t))$ from the first principle. Depending on the choice of the Hilbert space weight function $\rho$, the EMZ equations \eqref{gle_full}-\eqref{gle_projected} yield evolution equations for different dynamical quantities. When considering SDE \eqref{eqn:sde} in the context of statistical physics, the most common setting of $\rho$ is $\rho=\rho_0=\rho_{S}$, where $\rho_0=\rho_0(\bm x)$ is the distribution of the random initial condition (see \eqref{eqn:sde}), and $\rho_S=\rho_S(\bm x)$ is the steady state distribution of the stochastic system. For such a case, GLE \eqref{gle_full} yields the full dynamics of the noise-averaged quantity $\mathbb{E}_{\bm \xi(t)}[\bm u(\bm x(t))|\bm x_0]$, which is a {\em stochastic process} since the initial condition $x_0\sim\rho_0$ is random. On the other hand, the projected GLE \eqref{gle_projected} yields the evolution equation of the steady state time-autocorrelation function of $\bm u(\bm x(t))$, which is defined as \cite{pavliotis2014stochastic,zhu2020hypoellipticity} 
\begin{equation}\label{C(t)_def}
\begin{aligned}
C_{ij}(t):=\langle u_i(\bm x(t)),u_j(\bm x(t))\rangle_{\rho_S}=\P e^{t\K}u_i(0)&=\mathbb{E}_{\bm x_0}[\mathbb{E}_{\bm \xi(t)}[u_i(t)u_j(0)|\bm x_0]]\\
&=\langle\M(t,0)u_i(0),u_j(0)\rangle_{\rho_0}
=\langle\M(t,0)u_i(0),u_j(0)\rangle_{\rho_S}.
\end{aligned}
\end{equation}
Using the projected EMZ equation \eqref{gle_projected} to derive the evolution equation for the time auto-correlation function \eqref{C(t)_def} is the main technical difference between our EMZ framework and the ones used in \cite{hudson2018coarse,espanol1995hydrodynamics}. As we will see in Section \ref{sec:app}, approximating this projected equation is the key step of our reduced-order modeling. Other projection operators such as the Zwanzig-type projection are also used in the literature \cite{hudson2018coarse} to derive {\em nonlinear} GLEs for reduced-order quantities. This, however, is not the main focus of the current paper.

Lastly, we emphasize that the GLEs for deterministic Hamiltonian systems in the Gibbs equilibrium state $\rho=\rho_{eq}=e^{-\beta H}/Z$ satisfy the second fluctuation-dissipation theorem:
\begin{align}
    K_{ij}(t)=\sum_{k=1}^MG_{jk}^{-1}\langle u_k(0),\L e^{\Q\L\Q}\Q\L u_i(0)\rangle_{\rho_{eq}}
   &=-\sum_{k=1}^MG_{jk}^{-1}\langle \Q\L u_k(0), e^{\Q\L\Q}\Q\L u_i(0)\rangle_{\rho_{eq}} \nonumber \\
   &=-\sum_{k=1}^MG_{jk}^{-1}\langle f_k(0), f_i(t)\rangle_{\rho_{eq}} \label{G2ndFDT}
\end{align}
becasue of the idempotence of the symmetric operator $\Q$ and the skew-adjointness of the Liouville operator $\L$ with respect to the inner product $\langle\cdot,\cdot\rangle_{\rho_{eq}}$. However, since the Kolmogorov backward operator $\K$ in the EMZ equation is not skew-adjoint, the second fluctuation-dissipation theorem of the form \eqref{G2ndFDT} is no longer valid and needs to be generalized. We refer to our recent work \cite{zhu2021hypoellipticity3} for a more detailed explorations in this regard.

\section{Applications of the hypoelliptic analysis for the EMZ equation}
\label{sec:Ana}
In the previous section, we demonstrated how the EMZ equation is derived from the evolution operator $e^{t\K}$ and the orthogonal $e^{t\Q\K\Q}$. In this section, we focus on the prior estimation of these two semigroups and apply the established analytical results to various physical models. To be consistent with the literature on the hypoelliptic analysis, we will use the negative of $\K$ and $\Q\K\Q$ as semigroup generators and write the semigroups appearing in EMZ equation \eqref{eqn:EMZ_full} as $e^{-t\K}$ and $e^{-t\Q\K\Q}$. Moreover, all estimates are obtained first in the ``flat'' Hilbert space $L^2(\R^{d})$ and then transformed back in weighted Hilbert space $L^2(\R^{d};\rho)$.
The relationship between $L^2(\R^{d})$, $L^2(\R^{d};\rho)$ and the operators defined therein can be summarized using the following commutative diagram:
\begin{equation*}
\begin{tikzpicture}
  \matrix (m) [matrix of math nodes,row sep=3em,column sep=4em, minimum width=2em]
  {
     L^2(\R^{d}) & L^2(\R^{d};\rho) \\
     L^2(\R^{d}) & L^2(\R^{d};\rho) \\};
  \path[-stealth]
    (m-1-1) edge node [left] {$\tilde{\P},\tilde{\K},\tilde{\Q}$} (m-2-1)
            edge node [above] {$\U$} (m-1-2)        
    (m-2-2.west|-m-2-1) edge node [below] {$\U^{-1}$}
             (m-2-1)
    (m-1-2) edge node [right] {$\P,\K,\Q$} (m-2-2)
            ;
\end{tikzpicture}
\end{equation*}
where $\U$ and its inverse $\U^{-1}$ are unitary transformations which will be specified later for differential stochastic models. More detailed explanations can be found in \cite{zhu2020hypoellipticity}.
Throughout this paper, we denote the standard $L^2(\R^d)$ norm as $\|\cdot\|$. The inner product in $L^2(\R^d;\rho)$ is defined as \eqref{ip} with the induced weighted norm $\|\cdot\|_{L^2_{\rho}}$ given by 
\begin{align*}
\|\cdot\|_{L^2_{\rho}}=\left[\frac{1}{\int\rho d\bm x}\int(\cdot)^2\rho d\bm x\right]^{\frac{1}{2}}.
\end{align*}
Unless otherwise stated we only consider scalar quantities of interest. The following theoretical results were proved in \cite{zhu2020hypoellipticity}:
\begin{prop}[Zhu and Venturi \cite{zhu2020hypoellipticity}]
\label{exp_est_K}
Assuming a Kolmogorov operator $\tK$ of the form \eqref{KI} is a maximal-accretive operator in $L^2(\R^d)$ which satisfies the hypoelliptic conditions listed in Theorem 1 \cite{zhu2020hypoellipticity}.  
If the spectrum
of $\tK$ in $L^2(\R^n)$ is such that $\sigma(\tK)\cap i\R=\{0\}$, then there exits positive constants $\alpha$ and $C=C(\alpha)$ such that
\begin{align}\label{K_estimation}
\| e^{-t\tK}\tilde u_0-\tilde \pi_0\tilde u_0\|\leq C e^{-\alpha t}\|\tilde u_0\|,
\end{align}
for $\tilde u_0\in L^2(\R^n)$ and all $t>0$, where $\tilde \pi_0$ is the spectral projection onto the kernel of $\tK$. Moreover, the $n$-th order derivatives of the semigroup $e^{-t\tK}$ satisfies 
\begin{align}\label{power_K}
\|e^{-t\tK}\tK^n\|\leq \left(B_{\tK}\left(\frac{t}{n}\right)+\|\tilde \pi_0\tK\|\right)^n, \quad 
B(t)=C e^{-\alpha t}\left[1+\frac{1}{t}+\cdots+\frac{1}{t^M}\right]
\end{align}
for some positive constant $C$ and $M$.
\end{prop}
In \cite{zhu2020hypoellipticity}, it is further shown that the similar semigroup estimates hold for orthogonal semigroup $e^{t\tQ\tK\tQ}$ if $\tP=\I-\tQ$ is finite-rank, symmetric projection operator such as Mori's projection. In particular, we have
\begin{prop}[Zhu and Venturi \cite{zhu2020hypoellipticity}]\label{exp_est_QKQ}
Assume that $\tK$ satisfies all conditions listed in 
Proposition \ref{exp_est_K}. If $\tP: L^2(\R^n)\rightarrow L^2(\R^n)$ is a symmetric, finite-rank projection operator, and the spectrum 
of $\tQ\tK\tQ$ in $L^2(\R^n)$ is such that $\sigma(\tQ\tK\tQ)\cap i\R=\{0\}$,  then there exits positive constants $\alpha_{\tQ}$ and  $C=C(\alpha_{\tQ})$ such that
\begin{align}\label{QKQ_estimation}
\|e^{-t\tQ\tK\tQ}\tilde u_0-\tilde \pi^{\tQ}_0\tilde u_0\|\leq 
Ce^{-\alpha_{\tQ} t}\| \tilde u_0\|
\end{align}
for all $ \tilde u_0\in L^2(\mathbb{R}^n)$ and $t>0$, 
where $\tilde \pi^{\tQ}_0$ is the spectral projection 
onto the kernel of $\tQ\tK\tQ$. Moreover, the $n$-th order derivatives of the semigroup $e^{-t\tQ\tK\tQ}$ satisfies 
\begin{align}\label{power_QKQ}
\|e^{-t\tQ\tK\tQ}(\tQ\tK\tQ)^n\|\leq \left(B_{\tQ}\left(\frac{t}{n}\right)+\|\tilde \pi^{\tQ}_0(\tQ\tK\tQ)\|\right)^n, \quad 
B_{\tQ}(t)=C e^{-\alpha_{\Q} t}\left[1+\frac{1}{t}+\cdots+\frac{1}{t^{M_{\Q}}}\right],
\end{align}
for some positive constant $C$ and $M_{\Q}$.
\end{prop}
The proof of Proposition \ref{exp_est_K}-\ref{exp_est_QKQ} mainly uses the spectrum estimate for operator $\tK$ and $\tQ\tK\tQ$ and the functional calculus. The analysis is rather technical and hence will not be repeated here. In the following subsections, we focus on applying these theoretical results to specific stochastic dynamical systems.
\subsection{Application to Langevin dynamics}
\label{sec:app_EQ}
Consider the Langevin dynamics of an interactive particle system, described by the following system of SDEs in $\R^{2d}$:
\begin{align}\label{eqn:LE}
\begin{dcases}
\frac{d\bm q}{dt }=\frac{1}{m}\bm p\\
\frac{d\bm p}{dt} =-\nabla V(\bm q)-\frac{\gamma}{m}\bm p+\sigma \bm \xi(t)
\end{dcases}.
\end{align}
In eqn \eqref{eqn:LE}, $m$ is the mass of each particle, 
$V(\bm q)$ is the interaction potential and 
$\bm \xi (t)$ is a $d$-dimensional Gaussian white 
noise process modeling the physical Brownian motion.    
The parameters $\gamma$ and $\sigma$ are linked by 
the fluctuation-dissipation relation 
$\sigma=(2\gamma/\beta)^{1/2}$, where $\beta$ 
is proportional to the inverse of the thermodynamic 
temperature. The (negative) Kolmogorov operator \eqref{KI} associated with 
the SDE \eqref{eqn:LE} is given by
\begin{align}\label{L:LE}
\K=-\frac{\bm p}{m}\cdot\nabla_{\bm q}+
\nabla_{\bm q}V(\bm q)\cdot\nabla_{\bm p}+
\gamma\left(\frac{\bm p}{m}\cdot\nabla_p-\frac{1}{\beta}\Delta_p\right),
\end{align}
where ``$\cdot$'' denotes the standard dot product. 
If the interaction potential $V(\bm q)$ is strictly positive 
at infinity and satisfies the weak ellipticity assumption (Hypothesis 1 in \cite{zhu2020hypoellipticity}), then the Langevin equation \eqref{eqn:LE} 
admits an unique invariant Gibbs distribution
given by $\rho_{eq}(\bm p,\bm q)=e^{-\beta \H}/Z$,
where $\H=\frac{\|\bm p\|_2^2}{2m}+V(\bm q)$
is the Hamiltonian and $Z$ is the partition function. In 
\cite{zhu2020hypoellipticity}, it is further proved that Proposition \ref{exp_est_K} holds for any $\tilde u_0\in L^2(\R^{2d})$ and $t>0$ with $\tilde \pi_0(\cdot)=\langle (\cdot),e^{-\beta\H/2}\rangle e^{-\beta\H/2}$. Now we choose $\rho_{eq}$ as the weight of Hilbert space $L^2(\R^{2d};\rho_{eq})$, then the $L^2$-estimation \eqref{K_estimation} can be unitarily transformed \cite{zhu2020hypoellipticity} into the semigroup estimate in $L^2(\R^{2d};\rho_{eq})$ as:
\begin{align}
\|e^{-t\K} u_0-\pi_0 u_0\|_{L^2_{eq}}
\leq Ce^{-\alpha t}\|u_0\|_{L^2_{eq}},\label{eqn:Le^tL}
\end{align}
where $\pi_0(\cdot)=\langle(\cdot)\rangle_{eq}=\mathbb{E}[(\cdot)]$. Similarly, for orthogonal semigroup $e^{t\Q\K\Q}$ we have:
\begin{align}
\|e^{-t\Q\K\Q}u_0-\pi^{\Q}_0u_0\|_{L^2_{eq}}
&\leq Ce^{-\alpha_{\Q} t}\| u_0\|_{L^2_{eq}}.\label{QKQ_estimation1}
\end{align}
Different from the estimate for $e^{-t\K}$, the explicit expression of the kernel projection operator $\pi_0^{\Q}$ depends on the specific form of $\P$. For Mori-type projection operator $\P$ we considered, if there exists unique observable set $\{w_j\}_{j=1}^m$ such that $\langle w_j, u_i\rangle_{eq}=0$ and $\K w_j=\K^* w_j=u_j$, then $\pi_0^{\Q}$ admits analytical form
\begin{align}\label{PiQ}
\pi^{\Q}_0(\cdot)=\pi_0(\cdot)+\P(\cdot)+\sum_{i=1}^m\langle(\cdot),w_i\rangle_{eq}w_i.
\end{align}
Otherwise $\pi^{\Q}_0(\cdot)=\pi_0+\P$. With semigroup estimates \eqref{eqn:Le^tL} and \eqref{QKQ_estimation1}, we can derive prior estimations for different observable statistics. 

\vspace{0.2cm}
{\em Equilibrium state}. The equilibrium Langevin dynamics was studied thoroughly in \cite{zhu2020hypoellipticity}. Here we only review the key estimation result while the derivation is omitted. If the initial condition of the Langevin dynamics is \eqref{eqn:LE} set to be $\rho_0=\rho(t=0)=\rho_{eq}$, then the system is in a statistical equilibrium state, the corresponding dynamics is called the {\em equilibirum Langevin dynamics}. For equilibrium system, the time autocorrelation function $C(t)$ of a scalar observable $u(\bm x(t))=u(\bm p(t),\bm q(t))$ is stationary quantity satisfying $C(t,s)=C(|t-s|,0)$. Following the definition \eqref{C(t)_def}, we have
\begin{align}\label{def:correlation}
 C(t):=\mathbb{E}_{\bm x(0)}[\mathbb{E}_{\bm \xi(t)} [u(t)u(0)|\bm x(0)]]=\langle e^{t\K}u_0,u_0\rangle_{eq}.
\end{align}
Using Cauchy-Schwarz inequality and the semigroup estimate \eqref{eqn:Le^tL}, for $u_0\in L^2(\R^{2d};\rho_{eq})$ it is easy to get the asymptotic estimate for $C(t)$:
\begin{align}\label{asym_C_FPU}
|C(t)-\langle u_0\rangle^2_{{eq}}|
&=|\langle e^{-t\K}u_0,u_0\rangle_{{eq}}-\langle u_0\rangle_{{eq}}^2|\nonumber\\
&= |\langle e^{-t\K}u_0-\langle u_0\rangle_{{eq}},u_0\rangle_{{eq}}|\nonumber\\
&\leq \|e^{-t\K} u_0-\langle u_0\rangle_{{eq}}\|_{L^2_{eq}}\|u_0\|_{L^2_{eq}}
\leq Ce^{-\alpha t}\|u_0\|^2_{L^2_{eq}}.
\end{align}
This implies the equilibrium correlation function $C(t)$ approaches to the equilibrium value $\langle u_0\rangle^2_{eq}=\mathbb{E}^2[u_0]$ exponentially fast. To get the EMZ equation for observable $u(t)$, we introduce Mori-type projection $\P=\langle\cdot,u_0\rangle_{eq}u_0$. Substituting this into EMZ equation \eqref{gle_full} and \eqref{gle_projected} yields:
\begin{align}
\frac{d}{dt} u(t)&=\Omega u(t)+\int_0^t K(t-s) u(s)ds+f(t),\label{eqn:gle_FPU_EMZ}\\
\frac{d}{dt}C(t)&=\Omega C(t)+\int_0^t K(t-s)C(s)ds.\label{eqn:gle_FPU_PEMZ}
\end{align}
Here we note again that $u(t)$ is the actually the white nosie-averaged quantity $\mathbb{E}_{\bm \xi(t)}[u(\bm x(t))|\bm x(0)]$ and $\Omega=\langle u_0,\K u_0\rangle_{eq}/\langle u^2_0\rangle_{eq}$. 
By using Cauchy-Schwarz inequality and the semigroup estimate \eqref{QKQ_estimation1}, we can get the exponential convergence estimate for the EMZ memory kernel $K(t)$ and the fluctuation force $f(t)$:
\begin{align}
\left|K(t)-\langle\K^*_{eq} u_0,\pi_0^{\Q}\Q\K u_0\rangle_{eq}\right|
&\leq Ce^{-\alpha_{\Q}t}\|\K^*_{eq}u_0\|_{L^2_{eq}}\|\Q\K u_0\|_{L^2_{eq}},\label{K(t)_Langevin}\\
\Bigl\|f(t)-\pi_0^{\Q}\Q\K u_0\Bigr\|_{L^2_{eq}}
&\leq Ce^{-\alpha_{\Q} t}\|\Q\K u_0\|_{L^2_{eq}},
\label{f(t)_Langevin}
\end{align}
where $\K^*_{eq}$ is the adjoint operator of $\K$ in $L^2(\R^{2d};\rho_{eq})$ and the specific form of the kernel projection operator $\pi_0^{\Q}$ depends on $\P$ and the observable $u_0$, as we explained in \eqref{PiQ}. 

\vspace{0.2cm}
{\em Nonequilibrium nonsteady state.} Semigroup estimate \eqref{eqn:Le^tL} can also be used to get prior estimates for nonequilibrium Langevin dynamics. If the initial condition of \eqref{eqn:LE} is set to be $\rho_0=\rho(t=0)\neq \rho_{eq}$, the system evolves from a nonequilibrium nonsteady state. We now study the dynamics of the nonequilibirum mean function $M(t)$ defined as: 
\begin{align*}
M(t):=\mathbb{E}_{\bm x(0)}[\mathbb{E}_{\bm \xi(t)}[u(t)|\bm x(0)]]=\langle e^{t\K}u_0\rangle_{\rho_0}.
\end{align*}
$M(t)$ encodes the statistical moment information for a scalar observable $u(\bm x(t))$. Using the Cauchy-Schwarz inequality, the substitution $\rho_0=\rho_0\sqrt{\rho_{eq}}/\sqrt{\rho_{eq}}$ and the estimate \eqref{eqn:Le^tL}, we obtain the asymptotic estimate for $M(t)$:  
\begin{align*}
|M(t)-\langle u_0\rangle_{{eq}}|&=|\langle e^{-t\K}u_0\rangle_{\rho_{0}}-\langle\langle u_0\rangle_{{eq}}\rangle_{\rho_0}|\\
&= |\langle e^{-t\K}u_0-\langle u_0\rangle_{{eq}}\rangle_{\rho_{0}}|\\
&\leq \|e^{-t\K} u_0-\langle u(0)\rangle_{{eq}}\|_{L^2_{eq}}\left\|\rho_0^2/\rho_{eq}\right\|\leq Ce^{-\alpha t}\|u_0\|^2_{L^2_{eq}}
\left\|\rho_0^2/\rho_{eq}\right\|.
\end{align*}
Different from the equilibrium case, the convergence of the nonequilibrium mean $M(t)$ requires the finiteness of the $L^2(\R^{2d})$ norm $\|\rho_0^2/\rho_{eq}\|$, which imposes additional constraint on the initial probability distribution $\rho_{0}$. For instance, if the initial probability density is set to be the Gibbs distribution $\rho_0=e^{-\beta\H/4}/Z_{\beta/4}$ at high temperature $T\propto 4/\beta$, we have $\|\rho^2/\rho_{eq}\|=+\infty$ therefore the above estimate is not sufficient to guarantee the exponential convergence of $M(t)$ towards the equilibrium value $\langle u_0\rangle_{eq}$. Similar conclusion can be obtained from the return to equilibrium estimate for the probability density function $\rho(t,\bm p,\bm q)$ (see \cite{nier2005hypoelliptic}, Section 6.5):
\begin{align*}
\int_{\R^{2d}}\left|\rho(t,\bm p,\bm q)-\frac{1}{Z}e^{-\beta\H}\right|^2e^{\beta\H}d\bm pd\bm q\leq Ce^{-2\alpha_1t}.
\end{align*}
The above estimate is a dual of \eqref{K_estimation} and holds only for $\rho_0=\rho(0,\bm p,\bm q)\in e^{-\beta\H/2}\S'(\R^{2d})$, where $\S'(\R^{2d})$ is the space of tempered distributions. Obviously when $\rho_0=e^{-\beta\H/4}/Z_{\beta/4}\notin e^{-\beta\H/2}\S'(\R^{2d})$, there is no theoretical guarantee that the marginal distribution $\rho_u(t)$ would converges to the equilibrium marginal distribution. 
\subsection{Application to a heat conduction model} \label{sec:app_NEQ}
Consider a chain of nearest-neighbor interacting anharmonic oscillators coupled to two heat baths at end of the chain. Without adding external forces, the chain dynamics is determined by the system Hamiltonian:
\begin{align*}
\H_S(\bm p,\bm q)=\sum_{i=0}^N\left(\frac{p_i^2}{2}+V_1(q_i)\right)+\sum_{i=1}^NV_2(q_i-q_{i-1}).
\end{align*}
Now we attach the boundary oscillators to two thermostats with temperature $T_L$ and $T_R$, then the dynamics of the resulting heat conduction model \cite{eckmann1999non,eckmann2000non,eckmann2003spectral} is described by the system of stochastic differential equations:
\begin{equation}\label{sde_heat_cond}
\begin{dcases}
\begin{aligned}
dq_i&=p_idt,\qquad &&i=0,\cdots N,\\
dp_0&=-V_1'(q_0)+V'_2(q_1-q_0)dt+r_Ldt\\
dp_N&=-V_1'(q_N)+V'_2(q_N-q_{N-1})dt+r_Rdt\\
dp_j&=-V_1'(q_j)dt-V'_2(q_j-q_{j-1})dt+V_2'(q_{j+1}-q_j)dt,\qquad &&j=1,\cdots N-1  \\
dr_L&=-\gamma_Lr_Ldt+\lambda_L^2\gamma_Lq_0dt-\lambda_L\sqrt{2\gamma_LT_L}\xi_L(t)dt\\
dr_R&=-\gamma_Rr_Rdt+\lambda_R^2\gamma_Rq_Ndt-\lambda_R\sqrt{2\gamma_RT_R}\xi_R(t)dt
\end{aligned}
\end{dcases}
\end{equation}
where $\lambda_L,\lambda_R$ are the coupling constants between the boundary oscillators and the heat bath. $\xi_L(t)$ and $\xi_R(t)$ are the standard Gaussian white noise. The Kolmogorov backward operator $\K$ corresponding to the system of SDEs \eqref{sde_heat_cond} is given by:
\begin{equation}\label{Kolmogorov}
\begin{aligned}
\K=&\lambda^2_L\gamma_LT_L\partial_{r_L}^2+\lambda^2_R\gamma_RT_R\partial_{r_R}^2-\gamma_L(r_L-\lambda_L^2q_0)\partial_{r_L}-\gamma_{R}(r_R-\lambda_R^2q_N)\partial_{r_R}\\&+r_L\partial_{p_0}+r_R\partial_{p_N}
+\sum_{i=0}^N(p_i\partial_{q_i}-V_1'(q_i)\partial_{p_i})-\sum_{i=1}^NV_2'(q_i-q_{i-1})(\partial_{p_i}-\partial_{p_{i-1}}).
\end{aligned}
\end{equation}

{\em Equilibrium state.} When $T_L=T_R=T$, the system admits an invariant probability density which is given by the extended Gibbs distribution $\rho_{eq}=e^{-\beta\G(\bm p,\bm q,\bm r)}/Z$, where $\beta=1/T$ and $\G(\bm p,\bm q,\bm r)$ is the effective energy corresponding to the chain+heat bath system, defined as 
\begin{align}\label{effective_energy}
\G(\bm p,\bm q,\bm r)=\H_S(\bm p,\bm q)+\frac{r_L^2}{2\lambda_L^2}+\frac{r_R^2}{2\lambda_R^2}-q_0r_L-q_Nr_{R}.
\end{align}
The analysis for the equilibrium heat conduction model is exactly the same as the one for the Langevin dynamics. For potential energy $V_1$  and $V_2$ satisfying suitable conditions listed in \cite{eckmann2003spectral}, Eckmann and Hairer proved that the spectrum of the transformed Kolomogorov operator $\tilde{\K}$ in $L^2(\R^{2N+4})$ is discrete and has the cusp-shape spectrum $\S_{\tK}$. Therefore according to Proposition  \ref{exp_est_K}, if $\tK$ has no purely imaginary eigenvalue in $L^2(\R^{2N+4})$, then we have the exponentially decay estimate for scalar observable $u(\bm x(t))$:
\begin{align}
\|e^{-t\K} u_0-\pi_0 u_0\|_{L^2_{eq}}
&\leq Ce^{-\alpha t}\|u_0\|_{L^2_{eq}},
\label{est_e^tK_heat}\\
\|e^{-t\Q\K\Q}u_0-\pi^{\Q}_0u_0\|_{L^2_{eq}}
&\leq Ce^{-\alpha_{\Q} t}\| u_0\|_{L^2_{eq}},\label{QKQ_estimation_heat}
\end{align}
where weighted Hilbert space $L^2_{eq}=L^2(\R^{2d};\rho_{eq})$. Following the procedure outlined in Section \ref{sec:app_EQ}, it is easy to obtain corresponding exponentially decaying estimates for the equilibrium correlation function $C(t)$, EMZ memory kernel $K(t)$ and the fluctuation force $f(t)$. For the sake of brevity, the derivation details are omitted.

{\em Nonquilibrium steady state.} When $T_L\neq T_R$, it is proved in \cite{eckmann2000non} that the system admits an unique invariant measure $\mu$. Its density  $\rho_{S}$ is an smooth function on $\R^{2N+4}$ and can be represented as 
\begin{align}\label{heat_prob}
\rho_S=\tilde{h}(\bm p,\bm q,\bm r)e^{-\beta_0\G(\bm p,\bm q,\bm r)}.
\end{align}
In \eqref{heat_prob}, $\beta_0<\min\{\beta_L,\beta_R\}$, $\tilde{h}(\bm p,\bm q,\bm r)\in \bigcap_{\gamma>0}L^2(\R^{2N+4}; \G^{2\gamma}(\bm p,\bm q,\bm r))$ is a function decays faster than any polynomial as $\|\bm x\|\rightarrow \infty$. $\rho_{S}$ characterises a {\em nonequilibrium steady state} of the system. General speaking, it is hard to get an explicit expression of $\tilde h(\bm p,\bm q,\bm r)$, hence of the probability density \eqref{heat_prob}. However, we can still use Gibbs form equilibrium probability density $e^{-\beta\G}/Z$ as a reference state to derive prior estimates. To this end, we consider a weighted Hilbert space $L^2(\R^{2N+4};\rho_{r})$, where $\rho_r=e^{-2\beta_0\G}/Z$ and $1/\beta_0=T_0>\max\{T_L,T_R\}$. For the nonequilibrium case, the spectrum estimate obtained by Eckmann et al in \cite{eckmann2003spectral} still hold, which implies the following exponentially decay estimate for scalar observable $u(\bm x(t))$:
\begin{align}
\|e^{-t\K} u_0-\pi_0 u_0\|_{L^2_{r}}
&\leq Ce^{-\alpha t}\|u_0\|_{L^2_{r}}.
\end{align}
At the steady state, the correlation function $C(t)$ is stationary which can be defined as \eqref{C(t)_def} if the initial condition of \eqref{sde_heat_cond} satisfies $\rho(0)=\rho_{S}$. Using Cauchy-Schwarz inequality and the formal expression of the steady state density \eqref{heat_prob}, we obtain 
\begin{align*}
|C(t)-\langle u_0\rangle^2_{\rho_{S}}|
&=|\langle e^{-t\K}u_0,u_0\rangle_{\rho_{S}}-\langle u_0\rangle_{\rho_{S}},u_0\rangle_{\rho_{S}}|\\
&= |\langle e^{-t\K}u_0-\langle u_0\rangle_{\rho_{S}},u_0\rangle_{\rho_{S}}|\\
&\leq \|e^{-t\K} u_0-\langle u_0\rangle_{\rho_{S}}\|_{L^2_{r}}\|\tilde h(\bm p,\bm q,\bm r)u_0\|
\leq Ce^{-\alpha t}\|\tilde h(\bm p,\bm q,\bm x)u_0\|\|u_0\|_{L^2_r}.
\end{align*}
Since $\tilde{h}(\bm p,\bm q,\bm r)\in \bigcap_{\gamma>0}L^2(\R^{2N+4}; \G^{2\gamma}(\bm p,\bm q,\bm r))$, for any observable $u(\bm x(t))\in \S'(\R^{2N+4})$, e.g. polynomial functions, we have $\|\tilde h(\bm p,\bm q,\bm x)u_0\|\|u_0\|_{L^2_r}<+\infty$. The above estimate implies the steady state correlation function $C(t)$ decays to $\langle u_0\rangle^2_{\rho_{S}}$ exponentially fast. 
We emphasize that all estimates in Section \ref{sec:Ana} can be readily generalized to the $N$-dimensional EMZ equation \eqref{gle_full} and \eqref{gle_projected} where the observable $\bm u(\bm x(t))$ is a $N$-dimensional vector \cite{zhu2020hypoellipticity}.

\section{Memory kernel parametrization and the reduced-order modelling} 
\label{sec:param_memory}
From the previous discussion, we see that the prior estimation for the EMZ equation memory kernel implies that $K(t)$ is bounded by an exponentially decaying function. However, it does not answer what $K(t)$ exactly is, which is an important problem for the application the EMZ equation. In this section, we turn to focus on the numerical approximation of the EMZ memory kernel. The main method we will consider is the series expansion approach. Over the years, various basis functions have been used to construct approximation schemes of the classical system Mori-Zwanzig memory kernel \cite{baczewski2013numerical,berkowitz1981memory,Li2015,Li2017,lei2016data,chu2017mori,zhu2018faber}, where the expansion coefficients (parameters) are obtained through first-principle or data-driven methods. We will show that for the EMZ equation corresponding to the SDEs, similar approaches can be used to parametrize the memory kernel. In particular, we will prove that many commonly used date-driven methods are convergent due to the regularity of the orthogonal flow.

\subsection{The first-principle method of parametrization}
\label{sec:first-p}
A first-principle method to approximate the memory kernel was considered in \cite{zhu2018faber,zhu2019generalized}. It is shown that a series expansion of the memory kernel can be derived exactly from the semigroup expansion of orthogonal semigroup $e^{t\Q\K\Q}$. Following the derivation given therein, we consider the series expansion of the orthogonal semigroup:
\begin{equation}
e^{t\Q\K\Q}=\sum_{n=0}^{\infty} g_n(t)\Phi_n\left(\Q\K\Q\right),
\label{generalS}
\end{equation}
where $\Phi_n\left(\Q\K\Q\right)$ is the $n$-th order polynomial function of operator $\Q\K\Q$ and $g_n(t)$ the corresponding temporal basis. The simplest choice is the Taylor expansion where  $\Phi_n$ and $g_n$ are $\Phi_n\left(\Q\K\Q\right)=(\Q\K\Q)^n$
and $g_n(t)=t^n/n!$. Other possible choice of $\Phi_n$ ($n=0,\dots,N$) can be, e.g. the Faber polynomials \cite{zhu2018faber}, and the corresponding $g_n(t)=e^{-at}J_{n}(bt)$, where $J_n(bt)$ is the Bessel function of the first kind. Semigroup expansion \eqref{generalS} leads to function series expansions of the memory kernel. For a one dimensional EMZ equation, a substitution of \eqref{generalS} into \eqref{SFD} leads to 
\begin{align}
K(t)=\sum_{n=0}^{\infty} g_n(t)\frac{\langle\K\Phi_n\left(\Q\K\Q\right)\Q\K u_0,u_0\rangle_{\rho}}{\langle u_0,u_0\rangle_{\rho}}=\sum_{n=0}^{\infty}k_ng_n(t),
\label{K_series_expansion1}
\end{align}
where $k_n$ is the $n$-th expansion coefficient which can be understood as the operator cumulant averaged with respect to the probability density $\rho$. Naturally, a truncation of the expansion series \eqref{K_series_expansion1} yields an approximation of the exact memory kernel. From a theoretical point of view, it is hard to prove the convergence of expansion \eqref{K_series_expansion1} for nonlinear SDEs due to unboundedness of the operator $\Q\K\Q$. However, the validity of this approximation method has been verified numerically for linear and nonlinear Hamiltonian system in the statistical equilibrium \cite{zhu2018faber,zhu2019generalized}.

\vspace{0.2cm}
{\em First-principle method to calculate $k_n$.} 
The first-principle method calculate $k_n$ via the evaluation of the operator cumulants in \eqref{K_series_expansion1}. This can be realized using a recursive scheme and the associated combinatorial algorithm introduced in \cite{zhu2019generalized}. The original method is developed for the MZ equation of deterministic Hamiltonian systems. However, it can be readily generalized to stochastic system EMZ equation with some slight modifications of the derivation.  Here we only briefly review the main idea of the algorithm and refer to \cite{zhu2019generalized} for detailed explanations. Without loss of generality, it is convenient to consider a one-dimensional Mori's projection:
\begin{align}\label{Mori_P_1}
\P f=\frac{\langle f,u_0\rangle_{\rho}}{\langle u_0,u_0\rangle_{\rho}},
\end{align}
and introduce the following notation
\begin{equation}
\mu_i = \frac{\langle \K(\Q\K)^{i-1} u_0,u_0\rangle_{\rho}}{\langle u_0,u_0\rangle_{\rho}}, \qquad \gamma_i=\frac{\langle \K^i u_0,u_0\rangle_{\rho}}{\langle u(0),u(0)\rangle_{\rho}}.
\label{mugam}
\end{equation}
Clearly, if we are given $\{\mu_1,\dots, \mu_{n+2}\}$, 
then we can easily compute $\{k_1,\dots,k_n\}$ in \eqref{K_series_expansion1}, therefore the $n$-th order approximation of the memory kernel $K(t)$ for any given 
polynomial function $\Phi_n$. For example, if 
$\Phi_n(\Q\K\Q)=(\Q\K\Q)^n$ then $k_{q}=\mu_{q+2}/q!$ 
($q=0,\dots,n$). Directly evaluating $\mu_i$ is a daunting task since it involves taking operator powers and averaging of operator $\Q\K\Q$ which is a integral-differential operator by definition. However, the following recursive formula indicates that $\mu_i$ can be constructed iteratively from $\gamma_i$:
\begin{align}
\mu_1=\gamma_1,\qquad 
\mu_2=\gamma_2-\mu_1\gamma_1,\qquad\cdots,\qquad 
\mu_{n}&=\gamma_{n}-\sum_{j=1}^{n-1} \mu_{n-j} \gamma_{j}.
\label{iterative1}
\end{align} 
The proof of \eqref{iterative1} is provided in \ref{App0:proof}. Recurrence relation \eqref{iterative1} shifts the problem of 
computing $\{\mu_1,\dots,\mu_{n}\}$ to the problem 
of evaluating the coefficients $\{\gamma_1,\dots,\gamma_{n}\}$
defined in \eqref{mugam}. This can be done iteratively using the enumerative combinatorial algorithm introduced in  \cite{zhu2019generalized}, with the Livouille operator $\L$ used therein replaced by the Kolomogorov operator $\K$. For the sake of brevity, we omit technical details which can be found in \cite{zhu2019generalized}. In \ref{App1:Exact_method}, we provide the derivation of the combinatrorial algorithm for the Langevin dynamics \eqref{eqn:LE} of the Fermi-Pasta-Ulam (FPU) chain. 

\subsection{The data-driven method of parametrization}
\label{sec:data-driven}
Different from the first-principle method, there are many established data-driven methods which can be used to parametrize the memory kernel. Generally speaking, these methods use data collected by simulating stochastic dynamics \eqref{eqn:sde} to approximate the expansion coefficients $k_n$. The expansion series can be formulated in the temporal space as well as the frequency space \cite{lei2016data}. In this section, we are only concerned with the time-domain expansion and use the following ansatz to approximate $K(t)$:
\begin{align}\label{K_data_expansion}
K(t)\approx\sum_{n-0}^N\hat k_n\phi_n(t), \quad \text{where}\quad \hat k_n=\frac{\langle K(t),\phi_n(t)\rangle_{\omega}}{\langle\phi_n(t),\phi_n(t)\rangle_{\omega}}.
\end{align}
In \eqref{K_data_expansion}, $\{\phi_n(t)\}$ is the basis function defined in some open interval $I\subset\R^+$. The common choice of which are the orthogonal functions in a weighted Hilbert space $L^2(I,\omega)$. Under this setting, \eqref{K_data_expansion} becomes a generalized Fourier series. Hence, we can apply established results in approximation theory, say \cite{funaro2008polynomial}, to prove the convergence of the series expansion \eqref{K_data_expansion} as $N\rightarrow\infty$. As 
a preparation, we first use the Cauchy-Schwartz inequality and semigroup estimate \eqref{power_QKQ} to obtain the upper bounds of the $n$-th order derivative
\footnote{The definition of the $n$-th order derivative of $K(t)$ is a rather technical problem. In \eqref{K_n-estimate}, it is formally expressed using the time derivative of $e^{t\Q\K\Q}$, i.e. $K^{(n)}(t)=\langle \K^*e^{t\Q\K\Q}(\Q\K\Q)^n\Q\K u_0, u_0\rangle_{\rho}$. Mathematically, $K^{(n)}(t)$ is actually a weak derivative defined via the Dunford functional integral:
\begin{align}\label{Kn_def}
    K^{(n)}(t):=\left\langle\int_{\partial U}\lambda^ne^{t\lambda}\K^*R(\lambda,\Q\K\Q)\Q\K u_0d\lambda\cdot u_0\right\rangle_{\rho},
\end{align}
where $R(\lambda,\Q\K\Q)=(\lambda-\Q\K\Q)^{-1}$ is the resolvent of operator $\Q\K\Q$ and $\partial U$ is the boundary of the cusp $U$ which contains the spectrum of $\Q\K\Q$. Note that the right hand side of \eqref{Kn_def} is a smooth function of $t$, hence differentiable to an arbitrary order. More details on the weak convergence of the functional integral can be found in \cite{nier2005hypoelliptic,zhu2020hypoellipticity}.}

of the memory kernel: 
\begin{align}
|K^{(n)}(t)|&=|\langle \K^*e^{t\Q\K\Q}(\Q\K\Q)^n\Q\K u_0, u_0\rangle_{\rho}|\nonumber\\
&=|\langle e^{t\Q\K\Q}(\Q\K\Q)^n \Q\K u_0,\K^*_{\rho} u_0\rangle_{\rho}|\nonumber\\
&\leq C\|\Q\K u_0\|_{L^2_{\rho}}\|\K_{\rho}^*u_0\|_{L^2_{\rho}}\left(B_{\Q}\left(\frac{t}{n}\right)+\|\pi_0^{\Q}(\Q\K\Q)\|\right)^n,
\quad t>0.\label{K_n-estimate}
\end{align}
According to the definition of $B_{\Q}(t)$ in \eqref{power_QKQ}, estimate \eqref{K_n-estimate} implies that $K^{(n)}(t)$ is bounded by a continuous function of time in domain $I=(T_1,T_2)$, where $0\leq T_1\leq T_2<+\infty$. Hence for suitable weight function $\omega$, we have  $K(t)|_{I}\in\cap_{k=1}^{\infty}H_{\omega}^k(I)$, where $K(t)|_{I}$ is the restriction of $K(t)$ in the open interval $I$ and $H_{\omega}^k(I)$ is the weighted Sobolev space defined in $I$. This regularity provides sufficient conditions for the convergence of expansion \eqref{K_data_expansion}. If $\{\phi_n(t)\}$ is chosen to be, say the shifted Jacobi-type polynomials defined in $I$, then according to Theorem 6.2.4 in \cite{funaro2008polynomial}, the following convergence estimate holds for any $0<m<N$:
\begin{align}\label{hatK_est}
\left\| K(t)|_{I}-\sum_{n=0}^N\hat k_n\phi_n(t)\right\|_{L^2_{\omega}(I)}\leq \frac{C}{N^m}\left\|(1-t^2)^{m/2}K^{(m)}(t)\right\|_{L^2_{\omega}(I)},\qquad t\in I=(T_1,T_2).
\end{align}
Since $K(t)$ is naturally defined in domain $I=(0,+\infty)$, we can also set $\{\phi_n(t)\}$ to be the standard Laguerre polynomial with the weight function $\omega=e^{-t/2}$. For fixed $n\in\mathbb{N}^{+}$, using \eqref{K_n-estimate} we can get 
\begin{align}
\lim_{t\rightarrow +\infty}|K^{(n)}(t)|
\leq C\|\Q\K u_0\|_{L^2_{\rho}}\|\K_{\rho}^*u_0\|_{L^2_{\rho}}\|\pi_0^{\Q}(\Q\K\Q)\|^n,
\end{align}
which yields $|K^{(m)}(t)|t^{m/2}\in L^2_{\omega}(I)$. According to Theorem 6.2.5 in \cite{funaro2008polynomial}, this leads to the following convergence estimate for any $0<m<N$:
\begin{align}\label{hatK_est1}
\left\| K(t)-\sum_{n=0}^N\hat k_n\phi_n(t)\right\|_{L^2_{\omega}(I)}\leq \frac{C}{(\sqrt{N})^m}\left\|t^{m/2}K^{(m)}(t)\right\|_{L^2_{\omega}(I)},\qquad t\in I=(0,+\infty).
\end{align}
The semigroup estimate \eqref{power_QKQ} we used in the above derivation holds in the uniform topology for any scalar observable $u\in L^2(\R^n;\rho)$. As a consequence, the convergence rate we obtained on \eqref{hatK_est} and \eqref{hatK_est1} are not optimal. But we already know the convergence is {\em spectral}, i.e. faster than any polynomials. As far as we are concerned, this is the first convergence result for data-driven methods used in the Mori-Zwanzig framework. We also note that error estimate \eqref{hatK_est} only implies the approximation of $K(t)$ within $(T_1,T_2)$ is accurate.  In order to maintain low extrapolation error, in applications we will use basis functions defined in $I=(0,+\infty)$ to approximate the memory kernel. 

\vspace{0.2cm}
{\em Data-driven method to calculate $k_n$.} A substitution of the truncated expansion \eqref{K_data_expansion} into the projected EMZ equation \eqref{eqn:EMZ_projected} leads to the approximation scheme for $\P u(t)$. Since for Mori's projection, we have $\P u(t)=C(t)$ according to the definition \eqref{C(t)_def}, the scheme reads: 
\begin{align*}
\frac{d}{dt}C(t)\approx\Omega C(t)+\sum_{n=0}^{N}k_n\int_0^t \phi_n(s)C(t-s)ds,
\end{align*}
where the stationary correlation function $C(t)$ can be constructed from Monte-Carlo (MC) simulation data of the numerical solution to SDE \eqref{eqn:sde}, and the expansion coefficients $k_n$ can be obtained by solving numerically the following regression problem: 
\begin{align}\label{regression_pro}
\min_{\{k_n\}_{n=1}^N}\left\|\frac{d}{dt}C(t)-\Omega C(t)-\sum_{n=0}^{N}k_n\int_0^t \phi_n(s)C(t-s)ds\right\|_{L^2_{\omega}(I)}.
\end{align}
In Section \ref{sec:app}, we will use the LASSO regression  \cite{tibshirani1996regression} to solve \eqref{regression_pro} and get the approximated parameter set $\{k_n\}_{n=1}^N$. When compared with the first-principle method, the data-driven method in general has wider range of applicability but also demands more computational power because it requires the MC simulation data of the full dynamics.  

\subsection{Reduced-order modeling}
With the memory kernel $K(t)$ obtained using the first-principle or the data-driven parametrization method, we can now work on the reduced-order modeling for any low-dimensional observables $u(\bm x(t))$ of the stochastic system. Under the Mori-type projection, one may see that the projected EMZ equation \eqref{gle_projected} and the full dynamics \eqref{gle_full} (with random initial condition $\rho_0=\rho_S$) shares the memory kernel $K(t)$. In order to build a reduced-order model (ROM) for $u(\bm x(t))$ using the EMZ equation \eqref{gle_full}, it therefore boils down to the approximation of the fluctuation force $f(t)$. 

In the Mori-Zwanzig framework, $f(t)$ is formally given by $e^{t\Q\K\Q}\Q\K u_0$ which is also a stochastic process since the initial condition $u_0$ is random. Due to the randomness, it is hard to use techniques such as the operator series expansion \eqref{generalS} to approximate $f(t)$. However, since $u(t)$ in the steady state is a stationary stochastic process, $f(t)$ is also stationary and one may use the truncated Karhunen-Lo\'eve (KL) expansion series to approximate it. Without loss of generality, we assume $\langle f(t)\rangle_{\rho}=0$, then the KL expansion for $f(t)$ can be written as:
\begin{align}\label{KL_expansion_f}
f(t)\simeq \sum_{k=1}^{K}\eta_k\sqrt{\lambda_k}e_k(t),
\end{align}
where $\{\eta_k\}_{k=1}^K$ are the random coefficients and $\{\lambda_k,e_k\}_{k=1}^K$ are,
respectively, eigenvalues and eigenfunctions of the homogeneous Fredholm integral equation
of the second kind:
\begin{align}\label{Fredholm_eqn}
\int_0^T\langle f(t),f(s)\rangle_{\rho}e_k(s)ds=\lambda_ke_k(t),\qquad t\in[0,T],
\end{align}
where $T$ is a certain numerical integration time and $\langle f(t),f(s)\rangle_{\rho}$ is the time autocorrelation function of $f(t)$. In this paper, we only consider a specific case which allows us to determine the random coefficients  $\{\eta_k\}_{k=1}^K$ and the correlation function $\langle f(t),f(s)\rangle_{\rho}$ uniquely. To this end, we assume that the observable $u(t)$ is a Gaussian process and satisfies the second fluctuation-dissipation theorem: $\langle f(t),f(s)\rangle_{\rho}=K(|t-s|)$. It can be further verified \cite{zhu2019generalized} that $f(t)$ is also a Gaussian processes and its KL expansion random coefficients $\{\eta_k\}_{k=1}^K$ are necessarily i.i.d Gaussian random variables satisfying $\langle\eta_i\eta_j\rangle=\delta_{ij}$. As a result, we obtain the following ROM for $u(t)$:
\begin{equation}\label{ROM}
\begin{aligned}
    \frac{d}{dt}u(t)&= \Omega u(t)+\int_0^tK(t-s)u(s)ds +f(t)\\
    &\approx\Omega u(t)+\sum_{n=0}^N\int_0^tk_n\phi_n(t-s)u(s)ds+\sum_{k=1}^K\sqrt{\lambda_k}\eta_ke_k(t).
\end{aligned}
\end{equation}
By sampling the random coefficients $\{\eta_k\}_{k=1}^K$ and then solving numerically \eqref{ROM} with a proper numerical integrator, we obtain a ensemble of sample trajectories which, in principle, would imitate the dynamics of $u(\bm x(t))$ in the steady state. In Section \ref{sec:app}, we will also calculate the statistics from these simulated sample trajectories and compare them with the exact ones obtained from the molecular dynamics (MD) simulations to assess the effectiveness of the ROM.    

The modeling of $f(t)$ is harder when the observable $u(t)$ is a non-Gaussian process. In fact, this is a topic which is worth independent investigations. Here we only note some developed methods to address this problem. Specifically, Chu and Li \cite{chu2017mori} used a multiplicative noise to approximate $f(t)$. Zhu and Venturi \cite{zhu2019generalized} introduced a sample-based, transformated KL expansion to approximate the fluctuation force. In our recent work \cite{zhu2021hypoellipticity3}, a modified Sakamoto-Graham algorithm were proposed to do the modelling. On the other hand, as we briefly mentioned at the end of Section \ref{sec:EMZE}, the second fluctuation-dissipation theorem is {\em not} generally valid for {\em stochastic} system observables. Further explorations reveal that there exists a generalized second fluctuation-dissipation theorem for stochastic systems which can be used in reduced-order modelling. We refer to \cite{zhu2021hypoellipticity3} for more technical details.
\section{Applications}
\label{sec:app}
\begin{figure}[t]
\centerline{\hspace{0.1cm}
${\beta=1}$\hspace{7cm}
${\beta=20}$
}
\centerline{
\includegraphics[height=5.6cm]{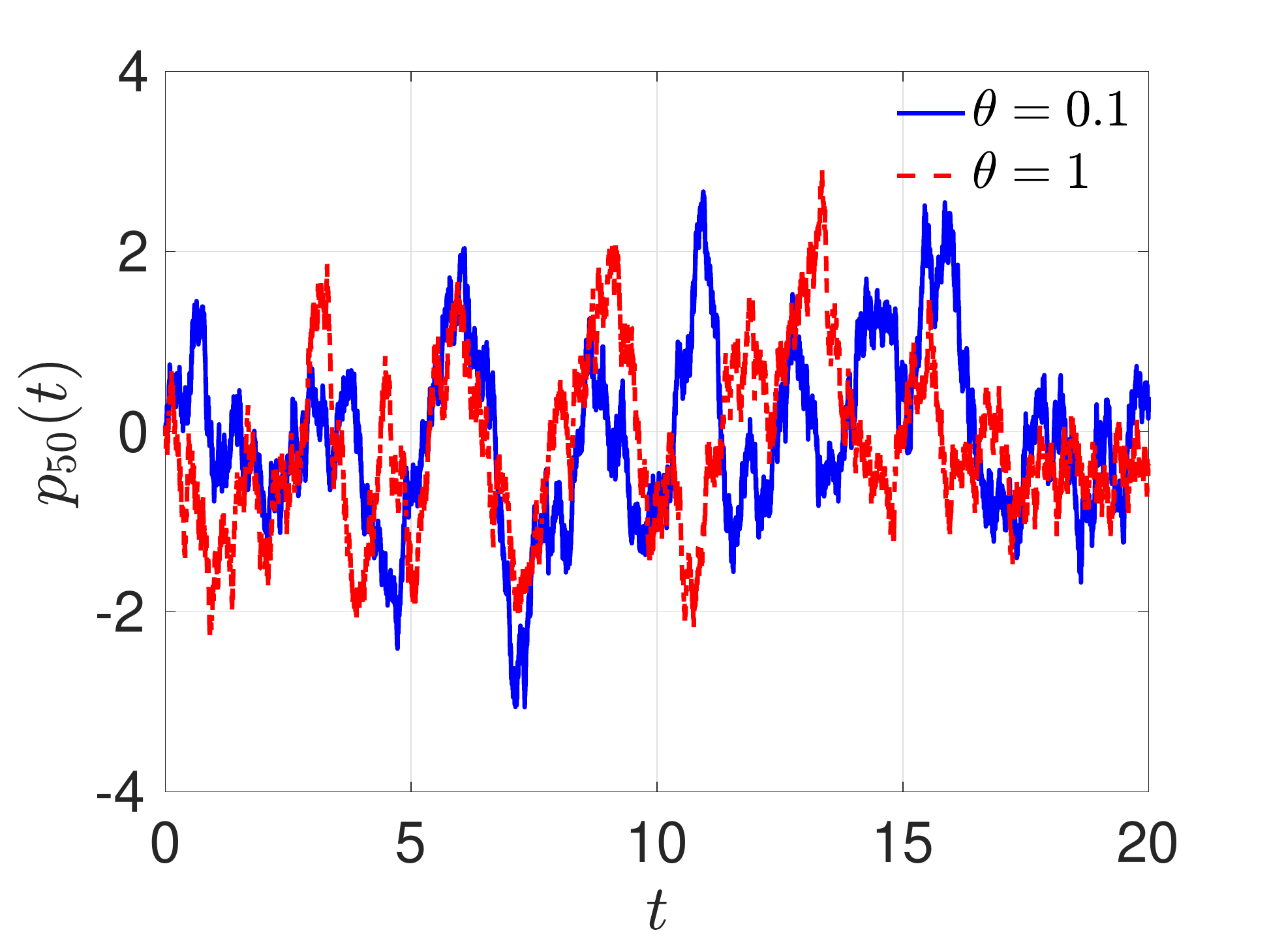}
\includegraphics[height=5.6cm]{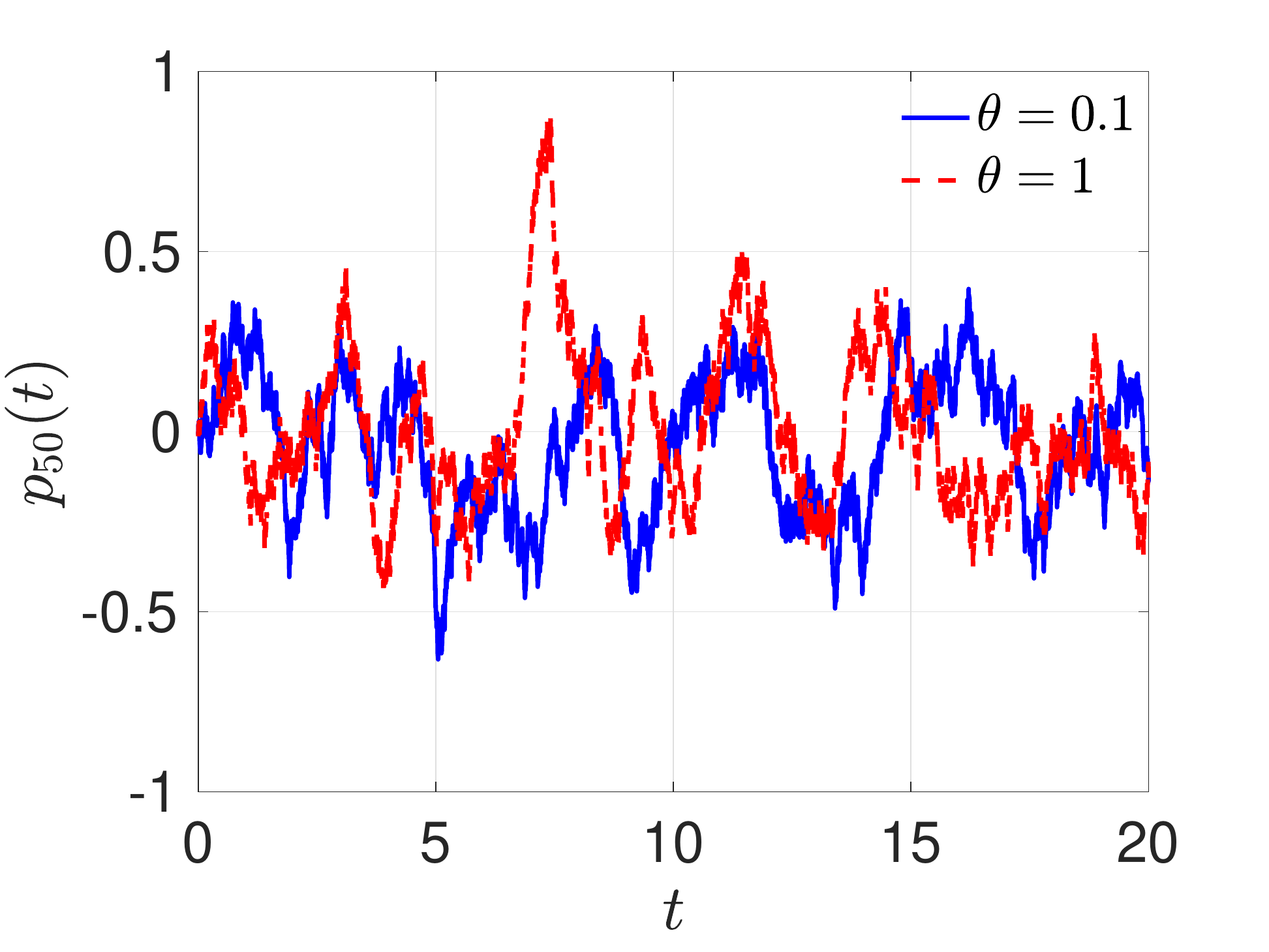}
}
\caption{Sample path of the tagged oscillator momentum $p_{50}(t)$. We display the result for the stochastic FPU system \eqref{eqn:FPU_Langein} with weak ($\theta=0.1$) and strong nonlinearity ($\theta=1$) at high ($\beta=1$) and low ($\beta=20$) temperature.}
\label{fig:Sample+path} 
\end{figure}
In this section, we will use the Langevin dynamics of a Fermi-Pasta-Ulam (FPU) chain model to numerically verify the theoretical results obtained in previous sections and validate the parametrization method of the EMZ memory kernel. To this end, we consider the Hamiltonian of the FPU chain:
\begin{align}\label{FPU_H}
\H(\bm p,\bm q)=\sum_{j=0}^{N-1}\frac{p_j^2}{2m}+\sum_{j=0}^{N-1}V(q_{j+1}-q_{j}),
\end{align}
where the potential energy is given by 
\begin{align}\label{FPU_potential}
V(q_{j+1}-q_{j})=\frac{\nu}{2}(q_{j+1}-q_{j})^2+\frac{\theta}{4}(q_{j+1}-q_{j})^4,
\end{align}
and $\{q_j, p_j\}$ are, respectively, the generalized 
coordinate and momentum of the $j$-th oscillator. In addition, the periodic boundary condition $q_0=q_{N}$ and $p_0=p_{N}$ is imposed, and the total number of oscillators is set to be $N=100$. For such a system, it is convenient to work on a new, non-canonical coordinate $\{\bm r,\bm p\}$ where $r_j=q_j-q_{j-1}$ is the distance between two neighboring oscillators. In the new coordinate, the Langevin dynamics \eqref{eqn:LE} for the stochastic FPU model is given by
\begin{align}\label{eqn:FPU_Langein}
\begin{dcases}
\frac{d}{dt}r_j&=\frac{1}{m}(p_j-p_{j-1}),\\
\frac{d}{dt}p_j&=\frac{\partial V(r_{j+1})}{\partial r_{j+1}}-
\frac{\partial V(r_{j})}{\partial r_{j}}-\frac{\gamma_j}{m}p_j+\sigma\xi(t).
\end{dcases}
\end{align}
The corresponding Kolmogorov backward operator is explicitly given by:
\begin{equation}\label{K_Langevin_Non}
\begin{aligned}
\K&=\L(\bm p,\bm r)+\S(\bm p)\\
&=\sum_{j=1}^{N-1}\left[
\left(\frac{\partial V(r_{j+1})}{\partial r_{j+1}}
-\frac{\partial V(r_j)}{\partial r_j}\right)\frac{\partial}{\partial p_j}+\frac{1}{m}(p_j-p_{j-1})\frac{\partial}{\partial r_j}\right]
-\sum_{j=1}^{N-1}\gamma_j\left(\frac{p_j}{m}\frac{\partial}{\partial p_j}-\frac{1}{\beta}\frac{\partial ^2}{\partial p_j^2}\right),
\end{aligned}
\end{equation}
where $\L(\bm p,\bm r)$ is the Liouville operator in the new coordinate $\{\bm r,\bm p\}$, $\S(\bm p)$ is an advection-diffusion operator involving $\bm p$. In Figure \ref{fig:Sample+path}, we display the sample paths of the momentum $p_{50}(t)$ of a tagged oscillator for SDE \eqref{eqn:FPU_Langein} with different parameters. 
\begin{figure}[t]
\centerline{
\includegraphics[height=5.6cm]{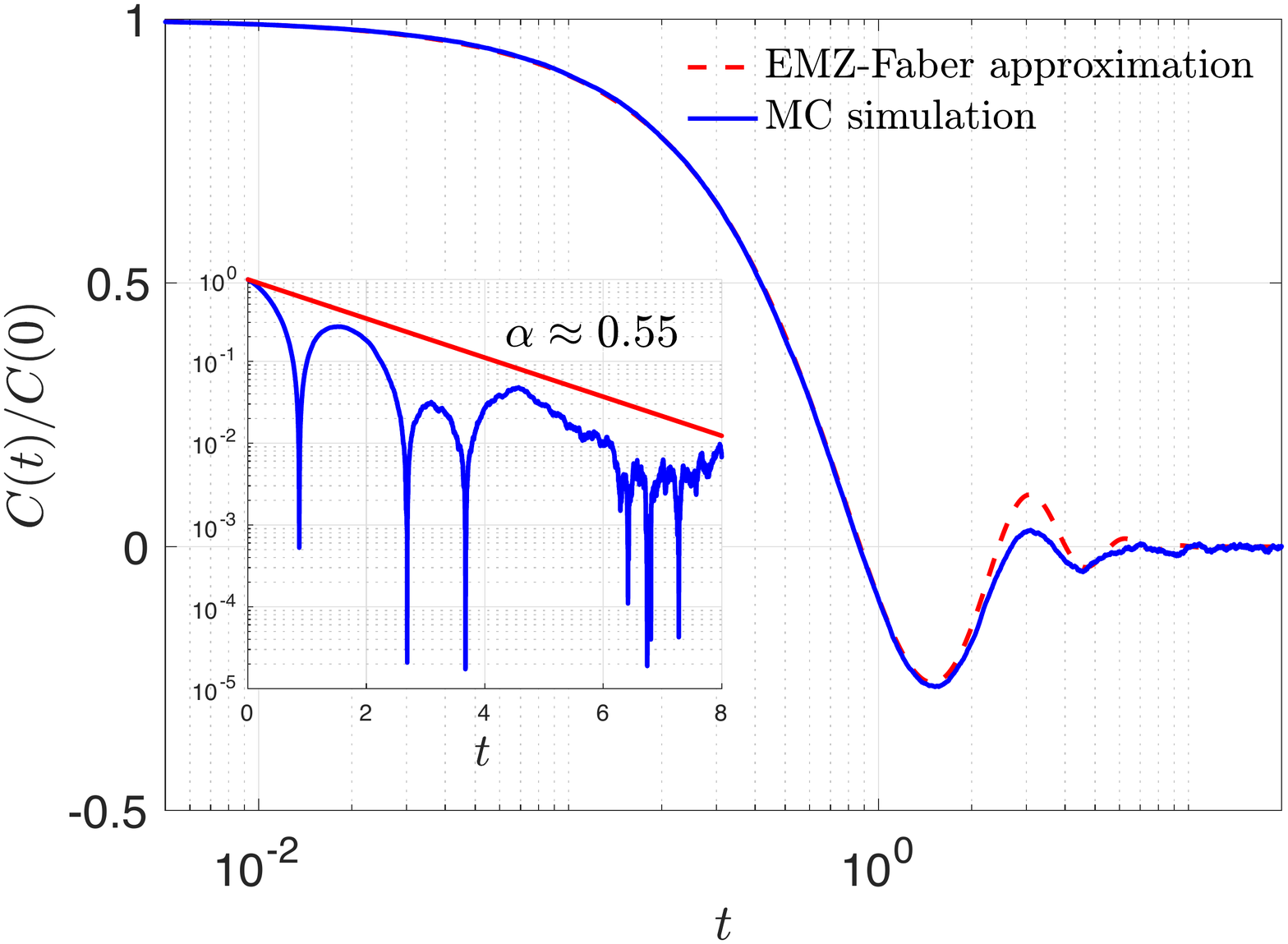}
\includegraphics[height=5.6cm]{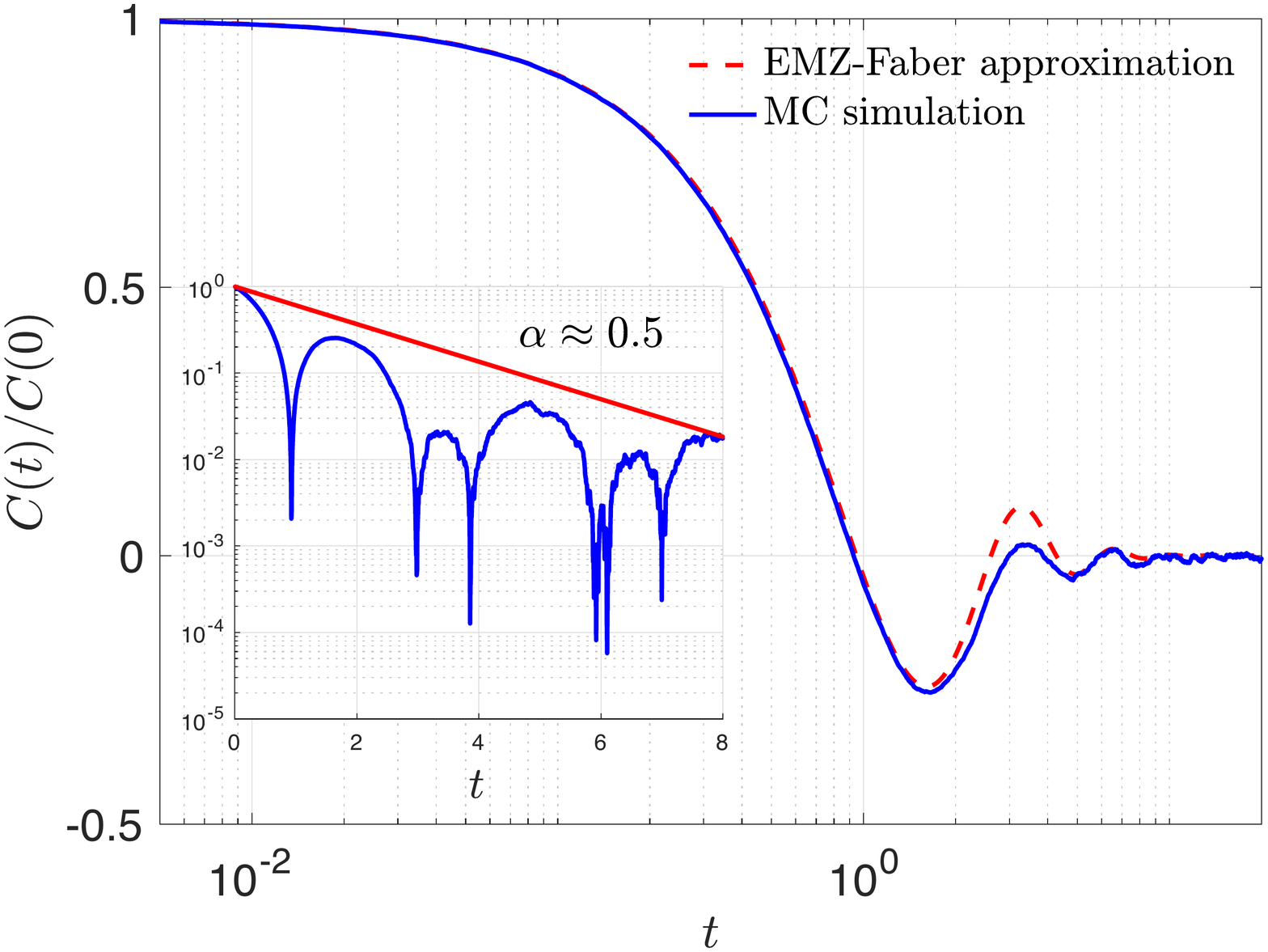}
}
\caption{Temporal 
auto-correlation function of the tagged oscillator momentum $p_j(t)$ for weakly nonlinear FPU system at different temperature $T\propto 1/\beta$. We compare results we obtained by calculating the EMZ memory from first principles using $14$-th order Faber polynomials with results from MC simulation ($10^6$ sample paths). In the subplots, we display $|C(t)/C(0)|$ and the exponentially decaying upper bound $ce^{-\alpha t}$ with an estimated decaying rate $\alpha$. }
\label{fig:C_FPU_weak} 
\end{figure}

\subsection{Memory kernel parametrization}
\vspace{0.2cm}
\noindent {\em Stochastic FPU chain with weak nonlinearity.}
We first consider the equilibrium dynamics of the FPU chain with weak nonlinearity. To this end, we set modeling parameter $\nu=m=1$, $\gamma_j=\gamma=1$ and $\theta=0.1$. The initial condition of \eqref{eqn:LE} is set to be $\bm x(0)\sim \rho_{eq}$ where $\rho_{eq}=e^{-\beta H}$ is the equilibrium Gibbs distribution. For the weakly nonlinear system, we aim to verify the following claims:

i) The observable statistics, in particular the auto-correlation function $C(t)$ and the corresponding memory kernel $K(t)$ defined in the projected EMZ \eqref{eqn:gle_FPU_PEMZ}, decays exponentially to its equilibrium value. 

ii) The first-principle method introduced in Section \ref{sec:first-p} yields accurate approximation to the memory kernel $K(t)$, therefore of $C(t)$.

\begin{figure}[t]
\centerline{\hspace{0.1cm}
${\beta=1}$\hspace{6.7cm}
${\beta=20}$
}
\centerline{
\includegraphics[height=5.6cm]{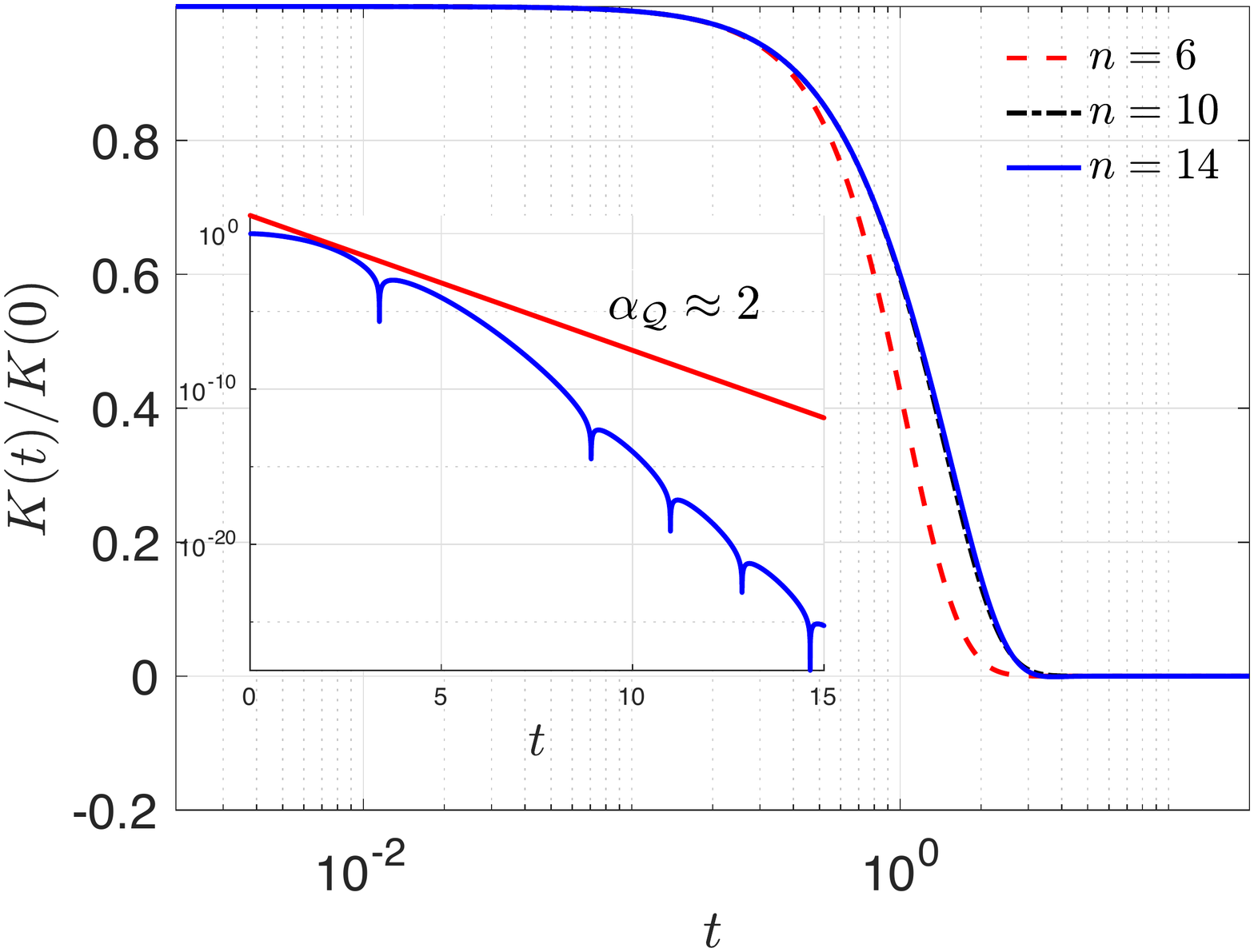}
\includegraphics[height=5.6cm]{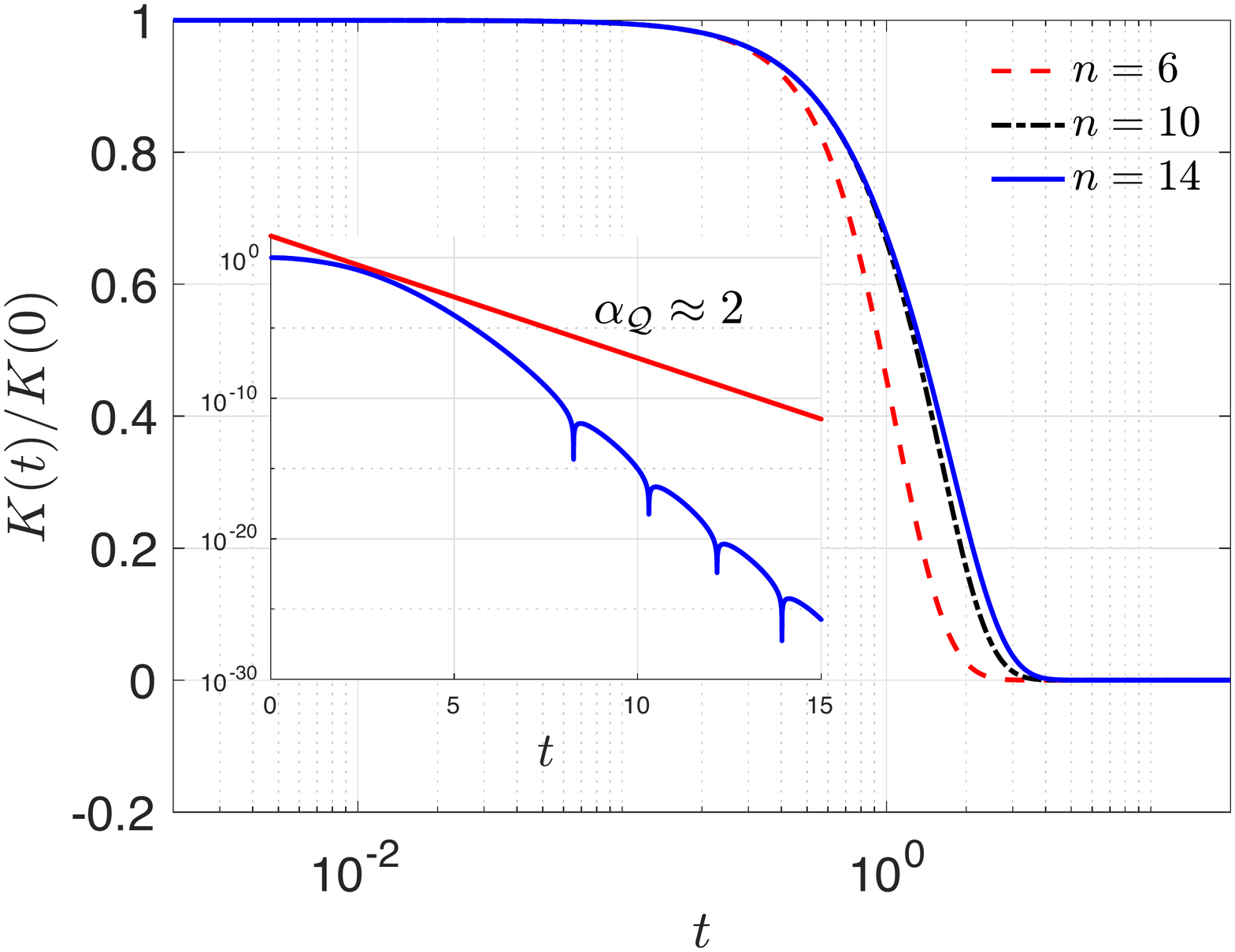}
}
\caption{Approximated EMZ memory kernel corresponding to the tagged particle momentum correlation function $C(t)$. The subplots display $|K(t)/K(0)|$ and the exponentially decaying upper bound $c_{\Q}e^{-\alpha_{\Q}t}$ with an estimated decaying rate $\alpha_{\Q}$. Other setting is same as Figure \ref{fig:C_FPU_weak}. }
\label{fig:K_FPU_weak} 
\end{figure}

For claim i), we note that the FPU potential energy defined as \eqref{FPU_potential} satisfies the weak ellipticity condition (Hypothesis 1) in \cite{zhu2020hypoellipticity}. Therefore the theoretical results in Section \ref{sec:app_EQ} hold for any polynomial-type observable. Now we choose the momentum  $p_j(t)$ of a tagged oscillator as quantity of interest and use Mori's projection $
\P(\cdot)=\langle(\cdot),p_j(0)\rangle_{eq}/\langle p_j^2(0)\rangle_{eq}$ to derive the projected EMZ equation \eqref{eqn:gle_FPU_PEMZ}. Some simple calculation implies $\Omega=-1$, hence the projected EMZ equation for the momentum yields the evolution equation for the time correlation function:
\begin{align}\label{eqn:gle_p_w}
\frac{dC(t)}{dt}=-C(t)+\int_0^tK(t-s)C(s)ds,
\end{align}
where $C(t)=\langle p_j(t),p_j(0)\rangle_{eq}$. According to estimate \eqref{asym_C_FPU}, the auto-correlation function $C(t)$ decays to the equilibrium value $\langle p_j(0)\rangle_{eq}^2=0$ exponentially fast with the rate $\alpha$. In the non-canonical coordinate $\{p_j, r_j\}$, we have $\K q_j=\K_{eq}^* q_j=p_j$. Moreover, since the {\em periodic} boundary condition is posed $q_j$ cannot be written as a function of $\{p_j, r_j\}$ (the linear transformation $\{q_j\}_{j=1}^N\rightarrow\{r_j\}_{j=1}^N$ is not invertible). Hence the kernel projection operator $\pi_0^{\Q}$ of $\Q\K\Q$ admits the explicit form:
\begin{align*}
\pi^{\Q}_0(\cdot)=\mathbb{E}[(\cdot)]+\P(\cdot),
\end{align*}
and the memory kernel estimate is given by \eqref{K(t)_Langevin}. Then we obtain $\langle\K^*_{eq}u_0,\pi_0^{\Q}\Q\K u_0\rangle_{eq}=0$ and the memory kernel estimate:
\begin{align*}
    |K(t)|\leq Ce^{-\alpha_{\Q}t}.
\end{align*}
where $C=C(p_i(0))$. In Figure \ref{fig:C_FPU_weak}, we plot the auto-correlation function $C(t)$ obtained by Monte-Carlo (MC) simulation ($10^5$ sample paths) for FPU systems with mild nonlinearities ($\theta=0.1$) at different temperatures ($\beta=1$ and $\beta =20$). The corresponding memory kernel $K(t)$ is shown in Figure \ref{fig:K_FPU_weak} which is obtained by the first-principle parametrization method. In both plots, we can see that $C(t)$ and $K(t)$ approaches to the predicted asymptotic $C(t=\infty)=0$ and $K(t=\infty)=0$ exponentially fast. 

For claim ii), we adopt the MZ-Faber expansion of $e^{-t\Q\K\Q}$ \cite{zhu2018faber} to approximate the memory kernel, where $e^{-at}J_n(bt)$ is the basis function. The simulation reult is displayed in Figure \ref{fig:C_FPU_weak}. It can be seen that the
MZ-Faber approximation of the EMZ memory kernel yields relatively accurate results for FPU systems with mild nonlinearties at both low ($\beta =20$) and high temperature ($\beta =1$).

\vspace{0.2cm}
\noindent {\em Stochastic FPU chain with strong nonlinearity.}
When the modeling parameter of the stochastic FPU chain is set to be $\nu=m=1$ and $\theta=1$, we get a strong nonlinear FPU chain. 
% We can similarly derive the projected EMZ equation for the time correlation function of the momentum of a tagged particle. The form of which will be the same as \eqref{eqn:gle_p_w}.
The first principle method introduced in Section \ref{sec:first-p} still can be applied here to approximate the EMZ memory kernel. However, large $\theta$ will lead to significant numerical instabilities at large $t$ when calculating $K(t)$ and $C(t)$ \cite{zhu2019generalized}. Hence in this paragraph, we will adopt the data-driven method to approximate the memory kernel. To this end, we use the standard Laguerre polynomial \cite{funaro2008polynomial} and Faber series \cite{zhu2018faber} to construct the data-driven approximation scheme of the EMZ memory kernel. In particular, the LASSO regression is used to solve \eqref{regression_pro} numerically to get the approximated parameter $\{k_n\}_{n=1}^N$. 
% By solving numerically the regression problem \eqref{regression_pro}, we can only get an {\em effective} memory kernel which may yield accurate predication of $C(t)$ or $\tilde u(t)$ while cannot be used to justify the exponential decaying estimate of $K(t)$. Considering this fact,
The data-driven method is used to verify the following claims:

iii) The auto-correlation function $C(t)$ defined in the projected EMZ \eqref{eqn:gle_FPU_PEMZ} decays exponentially to its equilibrium value. 

iv) The data-driven method introduced in Section \ref{sec:first-p} yields {\em effective} approximations to the memory kernel $K(t)$, therefore of $C(t)$.

\begin{figure}[t]
\centerline{\hspace{0cm}
${\beta=1}$\hspace{7.3cm}
${\beta=20}$
}
\centerline{
\includegraphics[height=5.7cm]{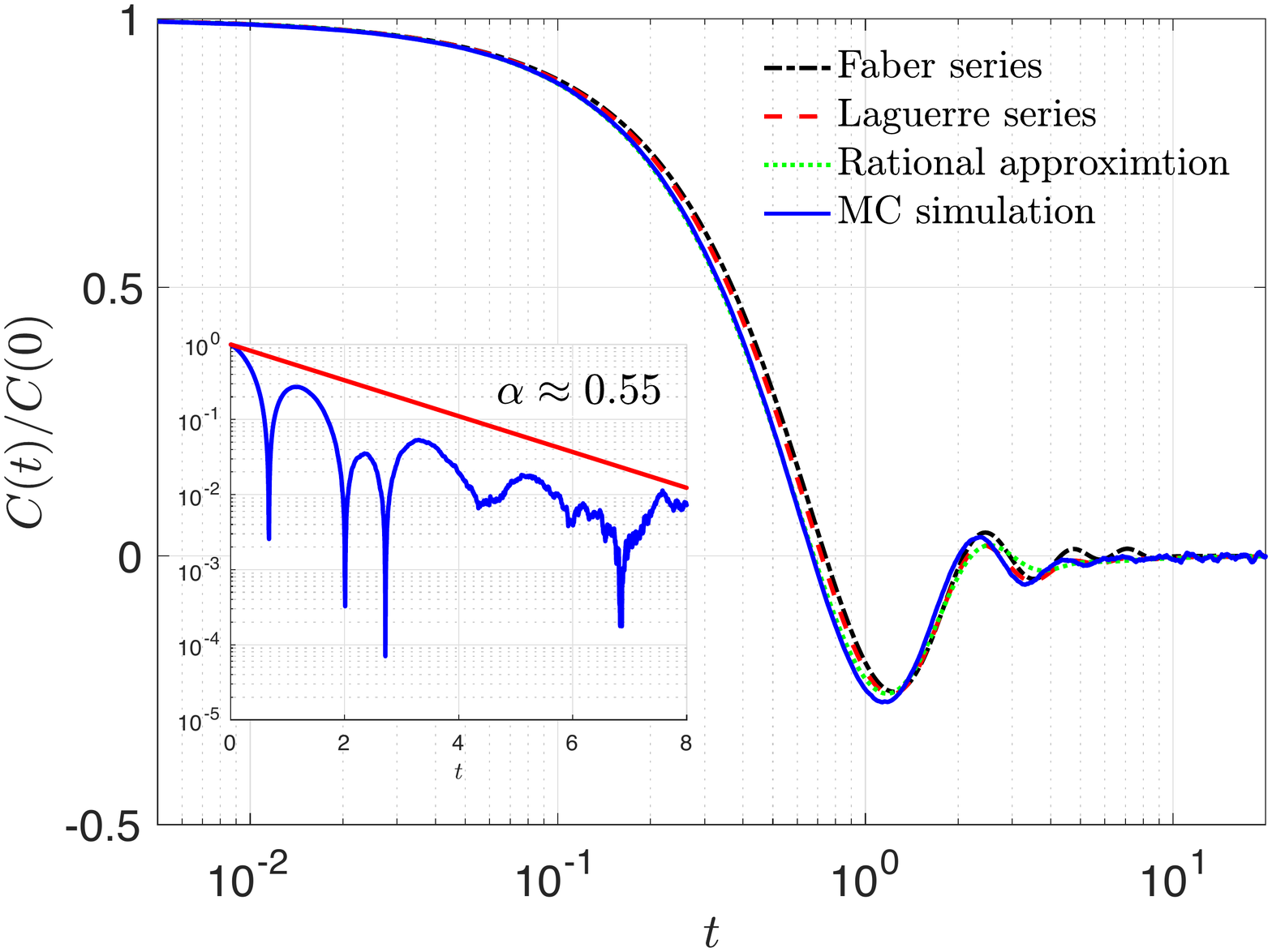}
\includegraphics[height=5.7cm]{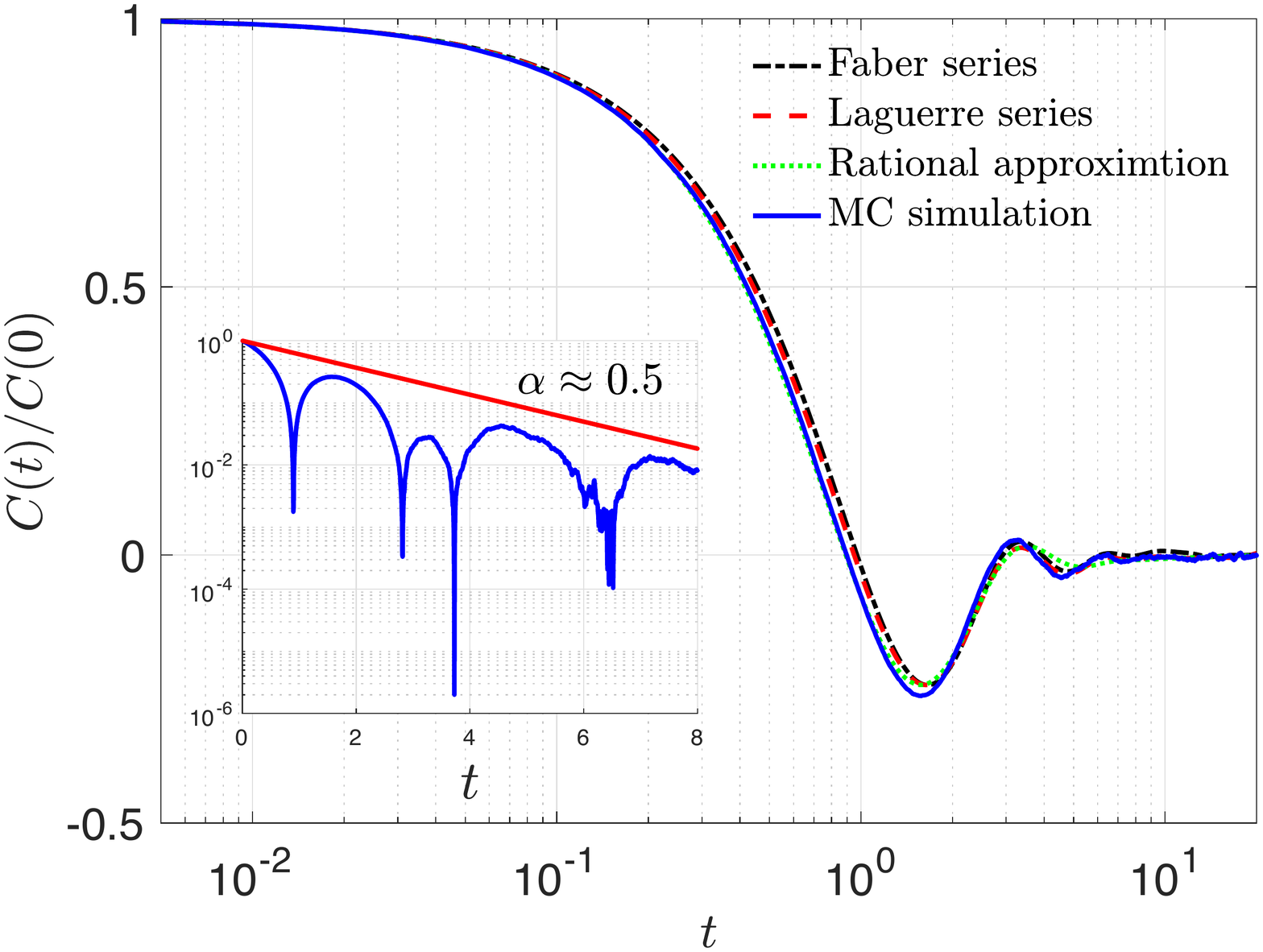}
}
\caption{Temporal 
auto-correlation function of the tagged oscillator momentum $p_j(t)$ for strongly nonlinear FPU system at different temperature $T\propto 1/\beta$. The MC simulation results ($10^6$ sample paths) of the correlation function are compared with the one obtained by the data-driven memory kernel using Faber series (20th order) and the standard Laguerre polynomials (20th order). In the subplots, we display $|C(t)/C(0)|$ and the exponentially decaying upper bound $ce^{-\alpha t}$ with an estimated decaying rate $\alpha$.}
\label{fig:C_FPU_strong} 
\end{figure}

To demonstrate iii), we use MC simulation ($10^5$ sample paths) to calculate the momentum auto-correlation function. It is shown in the subplots of Figure \ref{fig:C_FPU_strong} that $C(t)$ defined in the projected EMZ \eqref{eqn:gle_FPU_PEMZ} decays 0 exponentially fast. To validate iv), we adopt the Faber series and the standard Laguerre polynomials as the basis function to construct the data-driven approximation schemes for $K(t)$. These calculation results, along with the one obtained by the established rational approximation method \cite{lei2016data}, are presented in Figure \ref{fig:C_FPU_strong}. We can see that the data-driven method leads to accurate predication of $C(t)$.
\subsection{Reduced-order modeling}
\begin{figure}[t]
\centerline{
\includegraphics[height=4.cm]{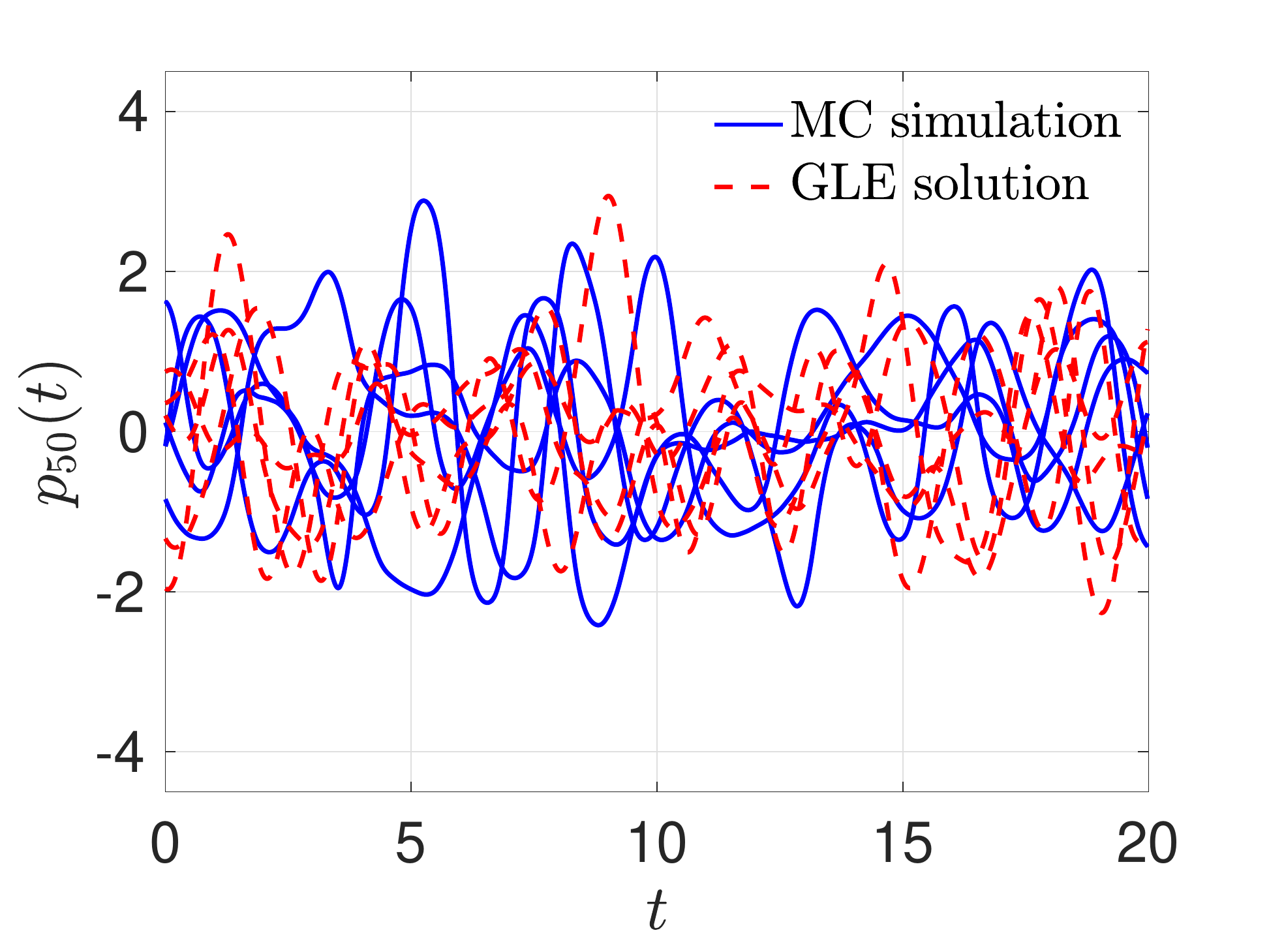}
\includegraphics[height=4.cm]{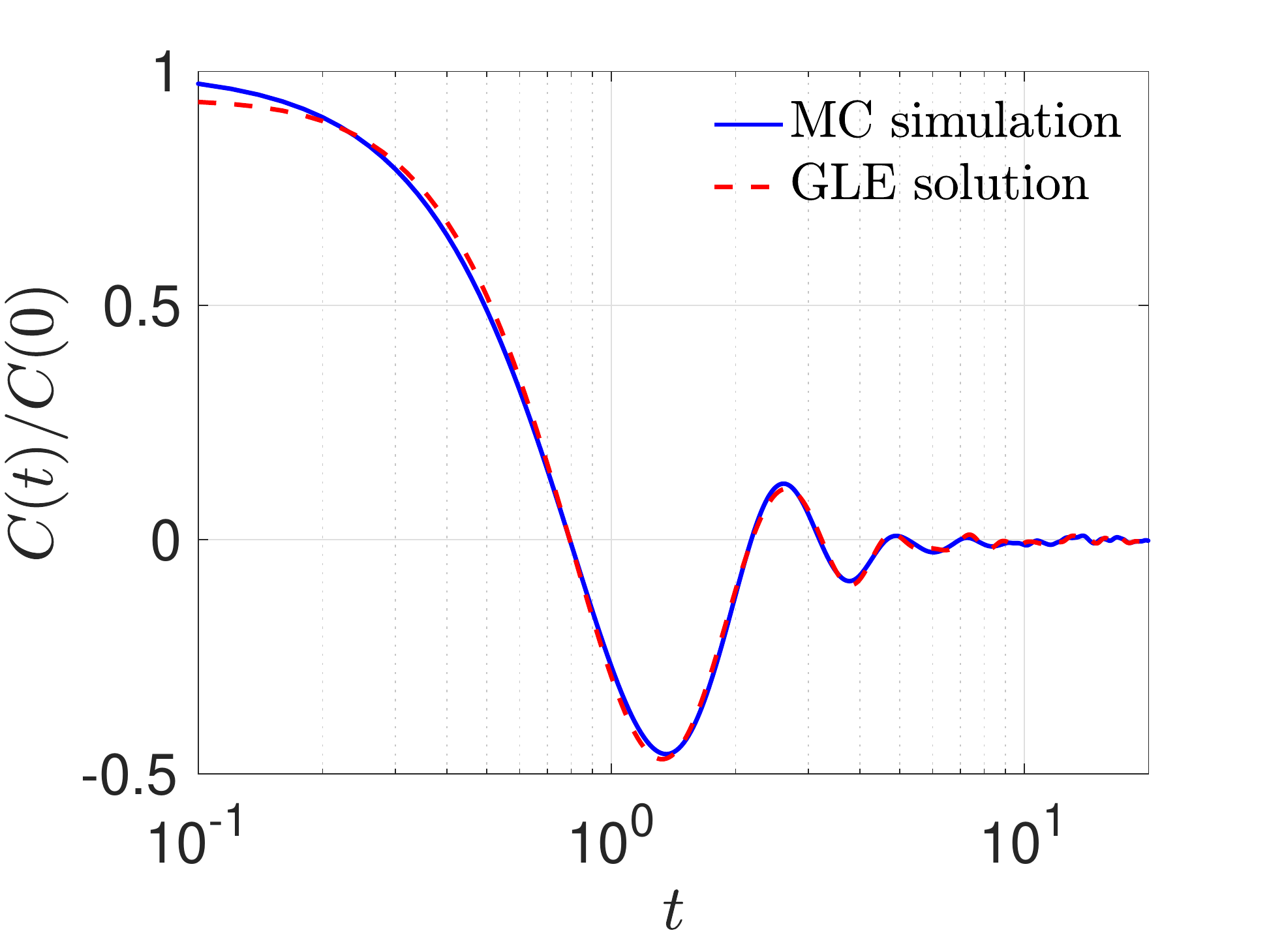}
\includegraphics[height=4.cm]{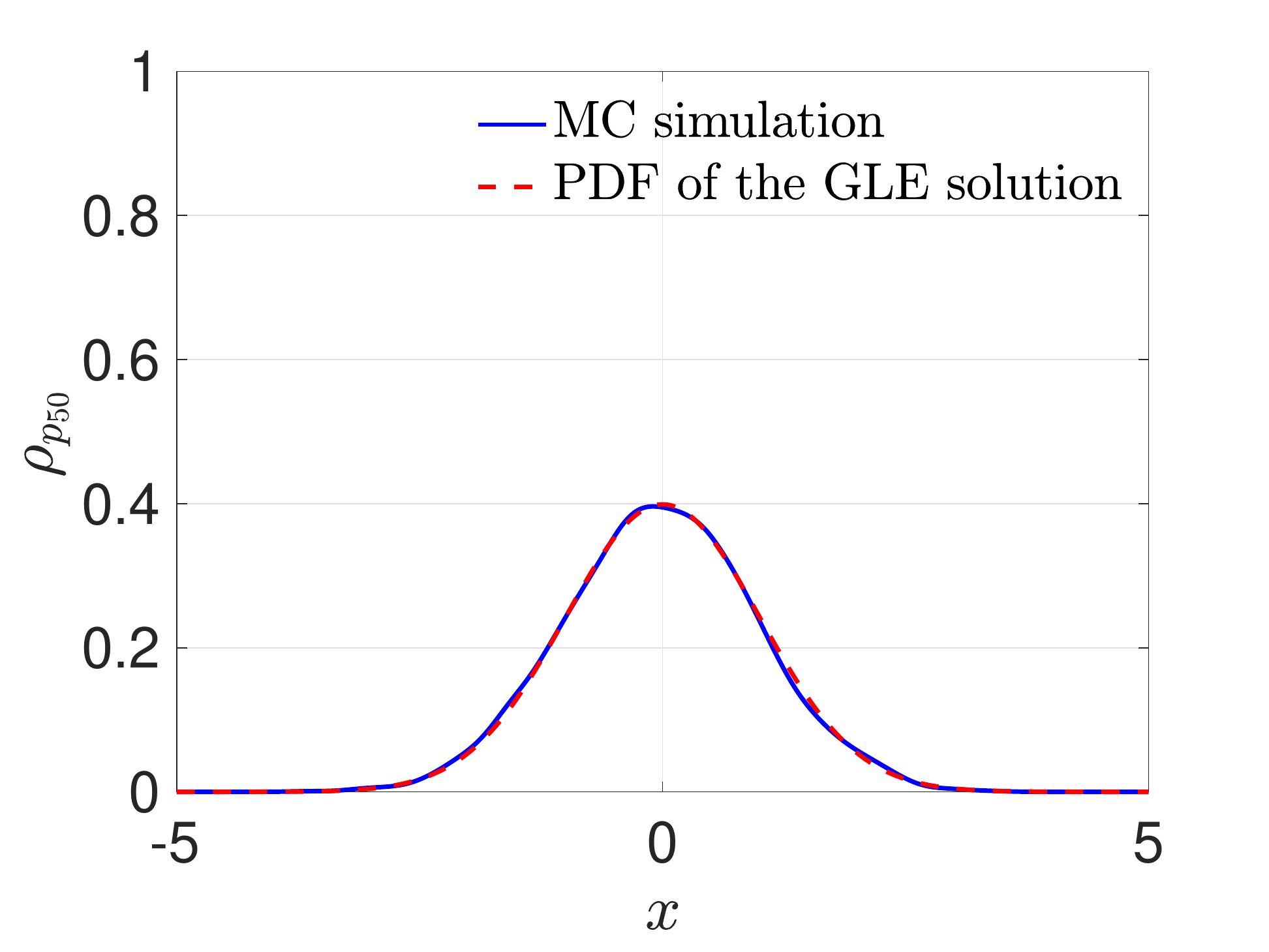}
}
\centerline{
\includegraphics[height=4.cm]{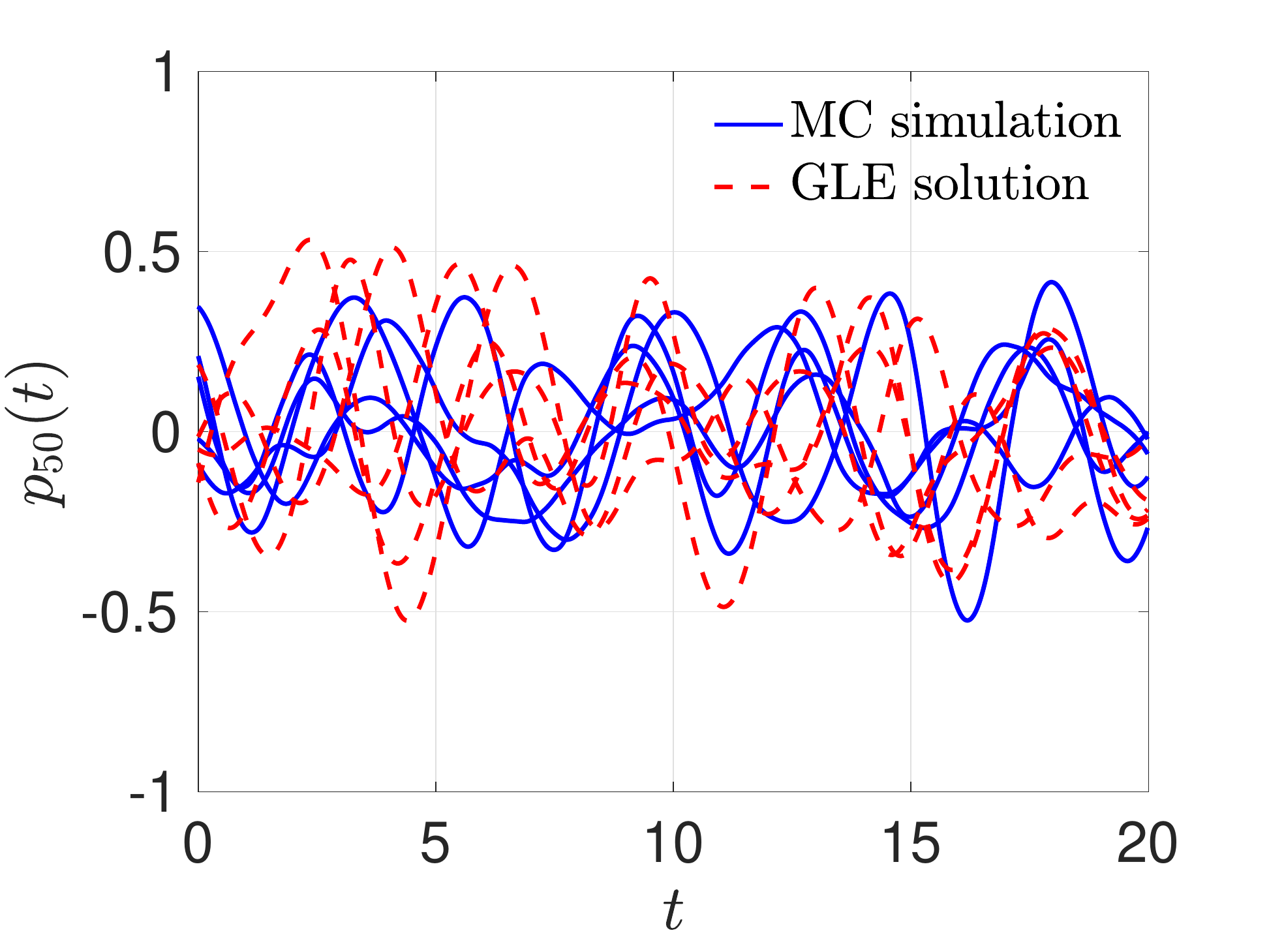}
\includegraphics[height=4.cm]{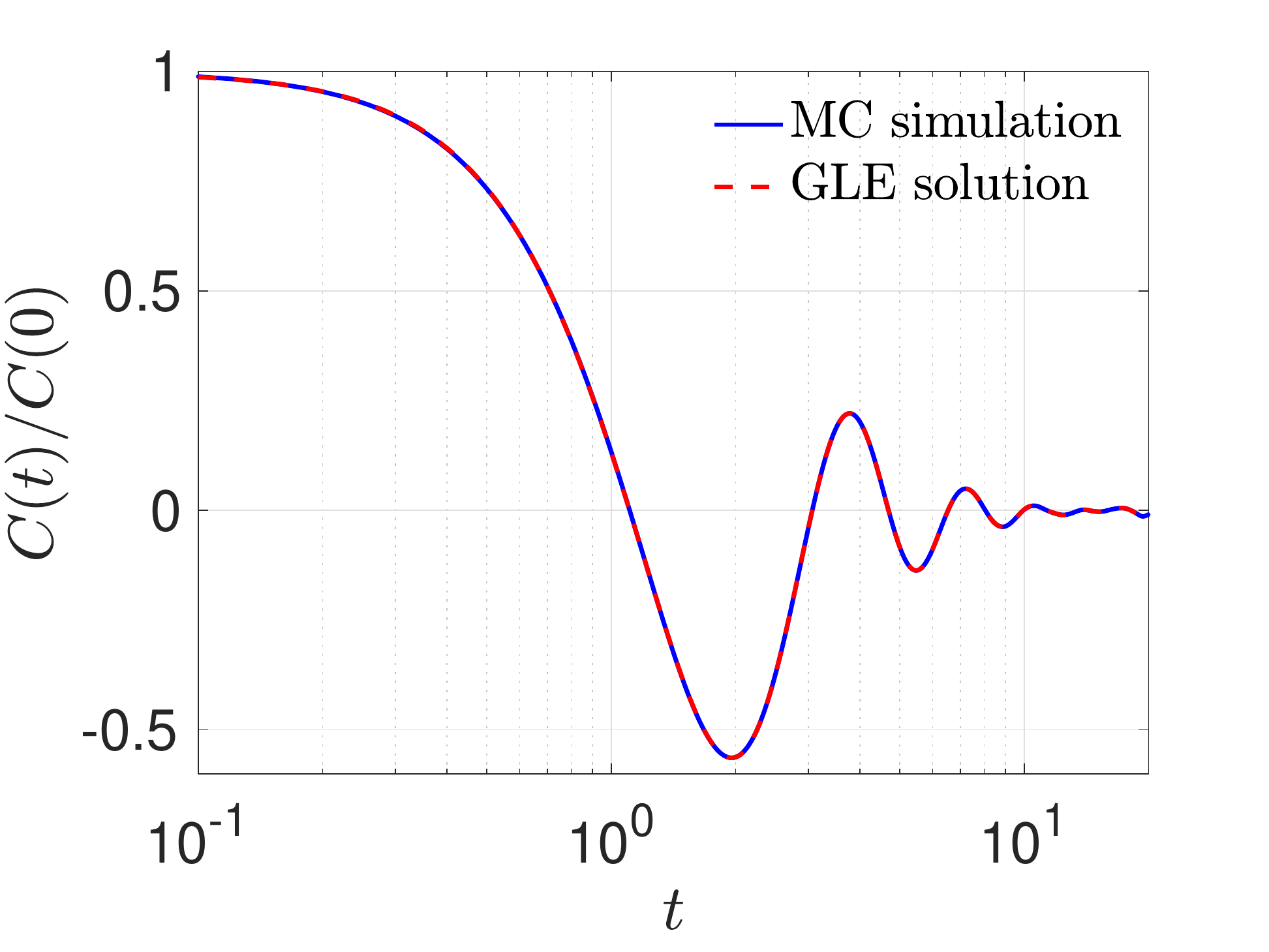}
\includegraphics[height=4.cm]{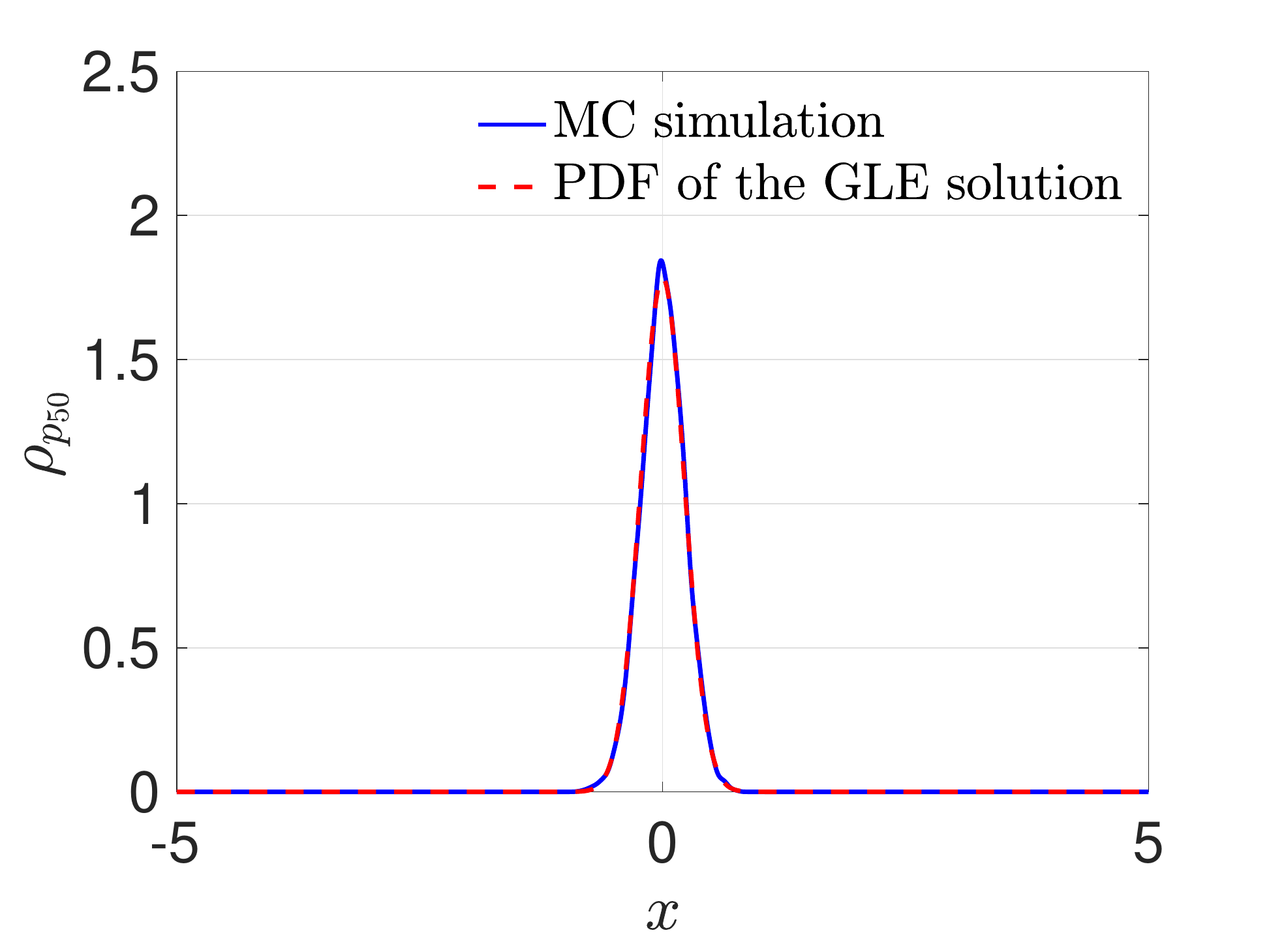}
}
\caption{Comparison of the dynamics of the particle momentum $p_{50}(t)$ generated by the MC simulation and the ROM \eqref{ROM_pj}. The displayed results are for a stochastic FPU system with strong nonlinearity ($\theta=1$) at high temperature $\beta=1$ (first row) and low temperature $\beta=20$ (second row). In the first column, we compare the simulated sample paths. The time autocorrelation functions $C(t)/C(0)$ (second column) are obtained by averaging a cluster of the sample trajectories. The third column compares the stationary distribution of the stochastic process $\rho_{p_{50}}$ which are obtained via kernel density estimations.}
\label{fig:p_Sample_path_compare} 
\end{figure}

In this subsection, we consider the equilibrium dynamics of the FPU chain with strong nonlinearity. However, we set the modeling parameters slightly different from that above with $\nu=m=
\theta=1$, $\gamma_{50}=1$ and $\gamma_j=0$ for $j\neq 50$\footnote{This parameter set is chosen such that the ROM model \eqref{ROM} for observable $p_{50}(t)$ satisfies the classical second FDT. The reason why this is the case is an interesting topic but out of the scope of the current paper. To this end, we refer to \cite{zhu2021hypoellipticity3} for detailed explanations.}. It is easy to verify that with this setting, the equilibrium Gibbs distribution $\rho_{eq}=e^{-\beta H}$ is still the stationary distribution of \eqref{eqn:FPU_Langein} with $\partial_t\rho_{eq}=\K^*\rho_{eq}=0$, where $\K^*$ is the adjoint of $\K$. Hence \eqref{eqn:FPU_Langein} yields an equilibrium dynamics.  Since in the equilibrium, $p_{50}(t)$ is obviously a Gaussian process, we can directly apply the ROM \eqref{ROM} to simulate the dynamics of $p_{50}(t)$. Specifically, we have: 
\begin{equation}\label{ROM_pj}
\begin{aligned}
    \frac{d}{dt}p_{50}(t)&\approx\Omega p_{50}(t)+\sum_{n=0}^N\int_0^tk_n\phi_n(t-s)p_{50}(s)ds+\sum_{k=1}^K\sqrt{\lambda_k}\eta_ke_k(t)\\
    &=\sum_{n=0}^N\int_0^tk_n\phi_n(t-s)p_{50}(s)ds+\sum_{k=1}^K\sqrt{\lambda_k}\eta_ke_k(t),
\end{aligned}
\end{equation}
where by simple calculations, we get $\Omega=0$. By sampling $\eta_k$ in \eqref{ROM_pj} and then solving it numerically using the 3rd-order Adams-Bashforth time integration scheme, we can get the solution of the ROM which can be regarded as a {\em realization} of $p_{50}(t)$ in the equilibrium. Figure \ref{fig:p_Sample_path_compare} compares the sample trajectories of the ROM and the path of $p_{50}(t)$ obtained by MC simulations. One can see that they are pretty much comparable with each other. We also calculate the time autocorrelation functions $C(t)/C(0)$ and the stationary marginal distributions $\rho_{p_{50}}$ of the stochastic process from the simulated sample paths. The correlation time of $p_{50}(t)$ is obviously longer than what obtained for the previous example. This difference is also reflected in the sample trajectories displayed in Figure \ref{fig:Sample+path} and Figure \ref{fig:p_Sample_path_compare} because the 
former ones are rougher. The obtained result indicates ROM \eqref{ROM_pj} imitates the dynamics $\rho_{p_{50}}$ in the equilibrium. We emphasize that the methodology also applies to nonequilibrium systems in the steady state.    
\section{Summary}
\label{sec:conclusion}
In this paper, we mainly focus on the application of the effective Mori-Zwanzig (EMZ) equation on the reduced-order modeling of stochastic systems. In particular, we showed that the semigroup estimates for $e^{-t\K}$ and $e^{-t\Q\K\Q}$ can be used to derive the exponentially decay upper bounds for various observable statistics associated with the EMZ equation, including the auto-correlation function $C(t)$, the EMZ memory kernel $K(t)$ and the fluctuation force. The results are presented for the Langevin dynamics of anharmonic oscillator chain and the heat conduction model in and out of statistical equilibrium. In addition, we introduced both the first-principle and data-driven methods to parametrize the EMZ memory kernel, and demonstrated that the regularity of $K(t)$ enables us to prove the convergence of frequently used data-driven approximation schemes. As far as we are concerned, this is the first theoretical convergence result regarding the approximation of the memory kernel. All these theoretical findings are verified numerically by simulating the Langevin dynamics for a Fermi-Pasta-Ulam (FPU) chain model. With the same example, we also proved the effectiveness of the numerical methods within their range of applicability. We conclude by emphasizing that analytical results obtained in this paper can be generalized and applied to the EMZ equation of other hypoelliptic stochastic systems. The numerical methodology we considered can be used to build effective reduced-order models for nonequilibrium systems in the steady state. 

\vspace{0.3cm}
\noindent 
{\bf Acknowledgements} 
Zhu's research was partially supported by the Air Force Office of Scientific Research (AFOSR) grant FA9550-16-586-1-0092. Lei's research was partially supported by NSF under grant DMS-2110981. The first author would like to thank Prof. D. Venturi and A. Iserles for stimulating discussions.
\appendix
\section{Proof of recurrence relation \eqref{iterative1}}\label{App0:proof}
The proof of the recurrence relation \eqref{iterative1} for equilibrium Hamiltonian system is given by Chu and Li in \cite{chu2017mori}. Here we provide a more general proof and show that such relation holds for {\em any} linear operator $\K$ and finite-rank projection operator $\P$. \eqref{iterative1} is a direct consequence of the following operator polynomial identity:
\begin{align}\label{sta}
\P\K(\Q\K)^{n}=\P\K^{(n+1)}-\sum_{i=1}^{n}\P\K(\Q\K)^{(i-1)}\P\K^{(n-i+1)},\qquad 0\leq n\in \mathbb{N}.
\end{align}
We can prove \eqref{sta} by induction. For $n=0$, we have identity $\P\K =\P\K$. Suppose for $n=k$, we have 
\begin{align*}
\P\K(\Q\K)^{k}=\P\K^{(k+1)}-\sum_{i=1}^{k}\P\K(\Q\K)^{i-1}\P\K^{(k-i+1)}.
\end{align*}
Then for  $n=k+1$, we have 
\begin{align*}
\P\K(\Q\K)^{k+1}&=\P\K(\Q\K)^{k}\K-\P\K(\Q\K)^{k} \P\K \\
&=\P\K^{(k+1)}\K-\sum_{i=1}^{k}\P\K(\Q\K)^{(i-1)}\P\K^{(k-i+1)}\K -\P\K(\Q\K)^{k} \P\K\\
&=\P\K^{(k+2)}-\sum_{i=1}^{k+1}\P\K(\Q\K)^{(i-1)}\P\K^{(k-i+2)}.
\end{align*}
By mathematical induction, the statement \eqref{sta} holds for all $0\leq n\in \mathbb{N}$. Applying operator identity \eqref{sta} on the observable $u(0)$ and then using the definition \eqref{Mori_P_1} and \eqref{mugam}, we can get the recurrence relation \eqref{iterative1}. 
For $M$-dimensional finite-rank projection \eqref{Mori_P}, using the same trick we get the following matrix form recurrence relation:
\begin{align}\label{iterative3}
\bm M_n=\bm \Gamma_n-\sum_{i=1}^{n-1}\bm\Gamma_{n-i}\bm M_i,
\end{align}
where $\bm M_n$, $\bm \Gamma_n$ are $M\times M$ dimensional matrix, defined as 
\begin{align*}
\P\K(\Q\K)^{(n-1)}\bm u(0)=\bm M_{n}\bm u(0),\qquad \P\K^n\bm u(0)=\bm \Gamma_{n}\bm u(0),
\end{align*}
and $\bm u(0)=[u_1(0),u_2(0),\cdots,u_M(0)]^T$ is the (initial) vector of quantities of interest and the range of $\P$ is $\text{Ran}(\P)=\text{Span}\{u_i(0)\}_{i=1}^M$. 
\section{First-principle algorithm to calculate $\gamma_n$ for stochastic FPU chain}
\label{App1:Exact_method}
The notaion used in this section follows exactly from \cite{zhu2019generalized}.  We first note that for a system with potential energy given by a polynomial function, the action of the $n$-th operator power $\K^n$ 
on a polynomial observable $u(\bm x)$ yields a polynomial function. Take $u(\bm x(t))=x_j$ as an example, this implies
\begin{align}\label{Ln}
\K^n x_j=\sum_{\bm b_i\in B^{(n)}}a^{(n)}_{\bm b_i} 
x_{k_{1}}^{m_{k_{1}}^{(i)}}\cdots x_{k_{r}}^{m_{k_{r}}^{(i)}}
\quad \Rightarrow
\quad
\K^{n+1} x_j=&\K \K^{n} x_j
=\sum_{\bm b_i\in B^{(n+1)}}a^{(n+1)}_{\bm b_i} 
x_{k_{1}}^{m_{k_{1}}^{(i)}}\cdots x_{k_{r}}^{m_{k_{r}}^{(i)}}.
\end{align}
where $\{a_{\bm b_i}^{(n)}\}$ are polynomial coefficients and  $\{m^{(i)}_{k_j}\}$ are polynomial exponents. At this point, it is convenient to define the set of polynomial exponents 
$B^{(n)}=\{\bm b_1,\bm b_2,\cdots \}$, the set polynomial 
coefficients $A^{(n)}=\{a^{(n)}_{\bm b_1},a^{(n)}_{\bm b_2},\cdots\}$, 
and the combined index set $\I^{(n)}=\{A^{(n)},B^{(n)}\}$.
Clearly, $\I^{(n)}$ identifies uniquely the 
polynomial \eqref{Ln}, i.e., there is a one-to-one correspondence between 
 $\I^{(n)}$ and $\K^n x_j$. If we can compute 
the mapping $\I^{(n)}\xrightarrow{\K} \I^{(n+1)}$, 
induced by the action of the Kolmogorov operator $\K$ to the polynomial 
\eqref{Ln} (represented by $\I^{(n)}$), then  
we can compute the {\em exact} series expansion 
of $\K^n x_j$ for arbitrary $n$. The whole process can be represented as  
\begin{align*}
u(\bm x) \rightarrow \K u(\bm x) 
\rightarrow \K^2u(\bm x) \rightarrow \cdots \rightarrow \K^nu(\bm x) 
\qquad\Longleftrightarrow\qquad
\I^{(0)}\xrightarrow{\K} \I^{(1)}
\xrightarrow{\K} \I^{(2)}\xrightarrow{\K}  \cdots\xrightarrow{\K}  \I^{(n)}
\end{align*}
where $\Longleftrightarrow$ represents the {\em translation} between the action of $\K$ on the observables and its action on the index set $\I^{(n)}$. It is left to determine the updating rule of $\I^{(n)}$ for the Langevin dynamics of the FPU chain. 
Suppose we are interested in the distance between 
the oscillators $j$ and $j-1$, i.e., in the polynomial 
observable $u(\bm p,\bm r)=r_j$. Using the formal definition of the Kolmogorov operator \eqref{K_Langevin_Non}, the action of 
$\K^n$ on $r_j$ can be explicitly written as  
\begin{align}\label{poly_form}
\K^nr_j=
\sum_{\bm b_{i}\in B^{(n)}}a_{\bm b_{i}}^{(n)}
r_{k_{1}}^{m_{k_{1}}^{(i)}}\cdots r_{k_{u}}^{m^{(i)}_{k_{u}}}
p_{l_{1}}^{s^{(i)}_{l_{1}}}\cdots p_{l_{v}}^{s^{(i)}_{l_{v}}},
\end{align}
where $\{k_{1},\dots,k_{u}\}$ and $\{l_{1},\dots,l_{v}\}$ are 
the relevant degrees of freedom for $\bm r$ and $\bm p$ 
at iteration $n$. We can explicitly compute the sets of such 
relevant degrees of freedom as  
\begin{align}\
K_r(n,j)=\left\{j-\left\lfloor\frac{n}{2}\right\rfloor,\dots,j+\left\lfloor\frac{n}{2}\right\rfloor\right\}\quad
L_p(n,j)=\left\{j-\left\lfloor\frac{n+1}{2}\right\rfloor,\dots,j+\left\lfloor\frac{n-1}{2}\right\rfloor\right\}.
\end{align}
The action of the Kolmogorov operator on each monomial
appearing in \eqref{poly_form} can be written as 
%%%
\begin{align}\label{L_iteration}
\K 
r_{k_{1}}^{m^{(i)}_{k_{u}}} r_{k_{u}}^{m^{(i)}_{k_{u}}} 
p_{l_{1}}^{s^{(i)}_{l_{1}}} \cdots p_{l_{v}}^{s^{(i)}_{l_{v}}} 
=
\sum_{v\in K_r(n,j)} \sum_{h\in L_p(n,j)} 
(\L_{r_{v}}+\L_{p_{h}}+\S_{p_{h}})
r_{k_{1}}^{m^{(i)}_{k_{1}}}\cdots r_{k_{u}}^{m^{(i)}_{k_{u}}}
p_{l_{1}}^{s^{(i)}_{l_{1}}}\cdots p_{l_{v}}^{s^{(i)}_{l_{v}}},
\end{align} 
%%%
where 
\begin{align*}
\L_{r_{v}}&=\frac{1}{m}(p_v-p_{v-1})\frac{\partial}{\partial r_v}\\
 \L_{p_{h}}&=\left[ \nu(r_{h+1}-r_{h})+\theta\left(r_{h+1}^3-r_{h}^3\right)\right]\frac{\partial}{\partial p_h}\\
\S_{p_{h}}&=\frac{\gamma p_h}{m}\frac{\partial}{\partial p_h}-\frac{\gamma}{\beta}\frac{\partial^2}{\partial p_h^2}.
\end{align*}
The action of $\L_{r_{v}}$, $\L_{p_{h}}$ and $\S_{p_h}$ on the monomial 
$r_{k_{1}}^{m^{(i)}_{k_{1}}}\cdots r_{k_{u}}^{m^{(i)}_{k_{u}}}
p_{l_{1}}^{s^{(i)}_{l_{1}}}\cdots _{l_{v}}^{s^{(i)}_{l_{v}}}$ 
can be explicitly computed.  This yields explicit linear 
maps of the polynomial exponents 
\begin{equation}
\bm b_i=[\bm m^{(i)},\bm s^{(i)}],\qquad 
\bm m^{(i)}=[m^{(i)}_{k_1}, \dots ,m^{(i)}_{k_u}],\qquad 
\bm s^{(i)}= [s^{(i)}_{l_1},\dots,s^{(i)}_{l_v}],
\end{equation}
and polynomial  coefficients 
$a_{\bm b_i}^{(n)}$.
With such maps available, we can transform the 
combined index set $\I^{(n)}$ (representing $\K^n r_j$)
to $\I^{(n+1)}$ (representing $\K^{n+1}r_j$). Specifically, we obtain
\begin{align*}
\I^{(n+1)}=\I^{(n+1)}_{\L_r} \biguplus \I^{(n+1)}_{\L_p}\biguplus  \I^{(n+1)}_{\S_p} ,
\end{align*}
where 
\begin{equation}\label{update_I^n}
\begin{aligned}
\I^{(n+1)}_{\L_r}
&=
\biguplus_{v\in K_r(n,j)}\biguplus_{i=1}^{\# B^{(n)}}
\biguplus_{k=0}^{1}\left\{m_v^{(i)}(-1)^ka_{\bm b_i}^{(n)},
[\bm m^{(i)}-\bm e_v,\bm s^{(i)}+\bm e_{v-k}]\right\},\\
 \I^{(n+1)}_{\S_p}
&=
\biguplus_{h\in L_p(n,j)}\biguplus_{i=1}^{\# B^{(n)}}
\left\{\{\gamma s_h^{(i)}a_{\bm b_i}^{(n)},-\frac{\gamma}{\beta} s_h^{(i)}(s_h^{(i)}-1)a_{\bm b_i}^{(n)}\},
\{
[\bm m^{(i)},\bm s^{(i)}], 
[\bm m^{(i)},\bm s^{(i)}-2\bm e_{h}]
\}
\right\},\\
\I^{(n+1)}_{\L_p}
&=\biguplus_{h\in L_p(n,j)}\biguplus_{i=1}^{\# B^{(n)}}
\biguplus_{k=0}^{1}\left\{\{s_{h}^{(i)}(-1)^{k+1}
\nu a_{\bm b_i}^{(n)},s_{h}^{(i)}(-1)^{k+1}
\gamma a_{\bm b_i}^{(n)} \},\right.\\
& \hspace{3.3cm}\left.\{[\bm m^{(i)}+\bm e_{h+k},\bm s^{(i)}-\bm e_{h}],
[\bm m^{(i)}+3\bm e_{h+k},\bm s^{(i)}-\bm e_{h}]\}\right\}.
\end{aligned}
\end{equation}
On the other hand, since $\K_{eq}^*=-\L(\bm p,\bm r)+\S(\bm p)$. It is easy to obtain the updating rule for the corresponding index set $\I^{*(n)}$ from the formal expression \eqref{update_I^n}. With these results available, we can immediately determine 
the coefficients $\gamma_j$ in \eqref{mugam} by averaging over the probability density $\rho_{eq}$ as 
\begin{equation}
\gamma_n=\frac{\langle\K^{n} r_j,r_j\rangle_{eq}}{\langle r_j,r_j\rangle_{eq}}=
\begin{dcases}
\frac{\langle\K^{\frac{n}{2}} r_j,\K_{eq}^{*\frac{n}{2}} r_j\rangle_{eq}}{\langle r_j,r_j\rangle_{eq}},\qquad &\text{$n$ is even},\\
\frac{\langle\K^{\frac{n+1}{2}} r_j,\K_{eq}^{*\frac{n-1}{2}} r_j\rangle_{eq}}{\langle r_j,r_j\rangle_{eq}}
,\qquad &\text{$n$ is odd}.
\end{dcases}
\label{gamma_n}
\end{equation}
Using formula \eqref{mugam}, \eqref{iterative1} and the exact expression of the polynomial $\Phi_n(\Q\K\Q)$, we can get the expansion coefficient $k_n$ in \eqref{K_series_expansion1}.

\bibliographystyle{plain}
\bibliography{hypo_app}
\end{document}